\documentclass[12pt]{article} %\input epsf.tex
\pdfoutput=1
\usepackage[utf8]{inputenc}
\usepackage[T1]{fontenc}
\usepackage{lmodern, amsmath, amssymb, slashed, graphicx, comment,adjustbox}
\usepackage[dvipsnames]{xcolor}
\usepackage[numbers, elide, compress]{natbib}
\usepackage[colorlinks=true, citecolor=blue, urlcolor=blue, linkcolor=blue, breaklinks=true, pdfpagelabels=false]{hyperref}

\usepackage[normalem]{ulem} 

\makeatletter\g@addto@macro\bfseries{\boldmath}\makeatother

\def\figureautorefname~#1\null{Fig.\,#1\null}

\def\equationautorefname~#1\null{Eq.\,(#1)\null}

\numberwithin{equation}{section}

\newcommand{\A}{\mathcal{A}}
\newcommand{\sym}{\mathcal{S}}

\newcommand{\bpm}{\begin{pmatrix}}
\newcommand{\epm}{\end{pmatrix}}

\newcommand{\la}{\langle}
\newcommand{\ra}{\rangle}

\usepackage{pifont}
\newcommand{\xmark}{\ding{55}}%

\begin{document}

\begin{flushright}
MITP/20-040
\end{flushright}

\vspace*{1.5cm}

\begin{center}

{\Large\bf
Sum Rules in the Standard Model Effective Field Theory from Helicity Amplitudes
\par}
\vspace{9mm}

{\bf Jiayin~Gu$\,^{a}$}~~and~~{\bf Lian-Tao Wang$\,^{b}$}\\ [4mm]
{\small\it
$^a$ PRISMA$^+$ Cluster of Excellence, Institut f\"ur Physik,\\
Johannes Gutenberg-Universit\"at, Staudingerweg 7, 55128 Mainz, Germany  \\[2mm]
$^b$ Department of Physics and Enrico Fermi Institute, University of Chicago, Chicago, IL 60637
\par}
\vspace{.5cm}
\centerline{\tt \small jiagu@uni-mainz.de, liantaow@uchicago.edu}

\end{center}

\vspace{1cm}

\begin{abstract}
The dispersion relation of an elastic 4-point amplitude in the forward direction leads to a {\it sum rule} that connects the low energy amplitude to the high energy observables. We perform a classification of these sum rules based on massless helicity amplitudes.  With this classification, we are able to systematically write down the sum rules for the dimension-6 operators of the Standard Model Effective Field Theory (SMEFT), some of which are absent in previous literatures.  These sum rules offer distinct insights on the relations between the operator coefficients in the EFT and the properties of the full theory that generates them. Their applicability goes beyond tree level, and in some cases can be used as a practical method of computing the one loop contributions to low energy observables.  They also provide an interesting perspective for understanding the custodial symmetries of the SM Higgs and fermion sectors. 
\end{abstract}

\newpage
{\small 
\tableofcontents}

\setcounter{footnote}{0}
%\pagebreak
%%%%%%%%%%%%%%%%%%%%%%%%%%%%%%%%%%%%%%%%%%%%%%%%%%%%%%%%%%%%%%%%%%

%%%%%%%%%%%%%%%%%%%%%%%%%%%%%%%%%%%%%%%%%%%%%%%%%%%%%%%%%%%%%%%%%%
\section{Introduction}
%%%%%%%%%%%%%%%%%%%%%%%%%%%%%%%%%%%%%%%%%%%%%%%%%%%%%%%%%%%%%%%%%%

Precision measurements and direct searches are the two 
cornerstones of high energy physics experiments. In terms of probing new physics beyond the Standard Model (SM), both approaches are important, and their complementarity is a crucial aspect in the planning of current and future collider experiments.  Effective Field Theories (EFT), with the assumption that new physics is heavy, offer a great way of parameterizing the results of precision measurements.
Direct search is, by definition, in the context of some model, which can be either a specific physics model or just a simplified one.  
In any given scenario, it is straightforward to connect these two approaches. 
The EFT at low energies can be obtained systematically from the full model via the matching procedure.  Yet, for a subset of observables related to elastic amplitudes, it is possible to relate the two approaches in a more general framework via dispersion relations.  Such relations, often denoted as {\it sum rules}, can be classified in terms of the energy expansion of the amplitudes analogous to the EFT operator expansion.  In previous works, the main focus has been at the level of dimension-8 operators for which the sum rules can be interpreted as positivity bounds on certain operator coefficients (or combinations of them) \cite{Adams:2006sv, Distler:2006if, Manohar:2008tc, Bellazzini:2015cra, Bellazzini:2016xrt, deRham:2017avq, deRham:2017zjm, Bellazzini:2017fep, deRham:2017xox, Chandrasekaran:2018qmx, deRham:2018qqo, Bellazzini:2018paj, Bi:2019phv, Bellazzini:2019bzh, Remmen:2019cyz, Remmen:2020vts, Wang:2020jxr, Zhang:2020jyn}.  
This remarkable finding suggests that the possible parameter space in an EFT is already constrained by the fundamental properties of quantum field theory, {\it i.e.}~unitarity, analyticity and locality.  
Unfortunately, testing these positivity bounds requires a decent determination on the dimension-8 operator  coefficients, which is difficult for current and near-future collider experiments. 
This is especially the case if one takes the validity of EFT into consideration~\cite{Contino:2016jqw}.  
For the dimension-6 operators which are phenomenologically more relevant, such positivity bounds could not be obtained in a model independent way.  Still, a number of interesting observations have been made.  For instance, the measured signs of certain dimension-6 operator coefficients could lead to strong predictions on the properties of heavy new particles~\cite{Low:2009di, Falkowski:2012vh, Bellazzini:2014waa}.

Scattering amplitudes provide the indispensable link between the full model and the EFT. 
The EFT needs to reproduce the amplitudes in the full model at the matching scale.  In a sum rule, the coefficients of the EFT can be directly related to observables in the full model.  This can be done by starting with an EFT Lagrangian, calculating the  low energy amplitudes in terms of the operator coefficients, and then using dispersion relations to connect it to the physics at higher energies.  Such procedures are straightforward and have been the standard practice in the literature.  Yet, in the spirit of the on-shell amplitude program (see {\it e.g.} Refs.~\cite{Elvang:2013cua, Dixon:2013uaa, Cheung:2017pzi} for recent reviews), it seems much more natural to directly treat amplitudes as a description of the EFT, which are equivalent to the Wilson coefficients of the higher dimensional operators in the Lagrangian.  Recent efforts have been made in parameterizing the Standard Model (SM) and its effective field theory (SMEFT) with on-shell amplitudes \cite{Arkani-Hamed:2017jhn,  Shadmi:2018xan, Ma:2019gtx, Aoude:2019tzn, Durieux:2019eor, Franken:2019wqr, Falkowski:2019zdo, Durieux:2019siw, Bachu:2019ehv}. While the conventional parameterization, obtained by adding higher dimensional operators to the SM Lagrangian~\cite{Buchmuller:1985jz, Grzadkowski:2010es}, still offers the most complete and practical description of the SMEFT, the on-shell approach does have certain advantages.  In particular, by working directly with the physical on-shell amplitudes, one is freed from the burden of operator redundancies and basis choices which can often obscure the physics picture.  
In this work, we take a similar approach by focusing on the massless elastic on-shell amplitudes in the SMEFT and use them to classify and study the sum rules.  A great advantage of massless amplitudes is that they can be characterized based on the helicities of the particles, and take particularly simple forms.   
Practically, the massless limit can be realized by considering measurements with energy ($E$) sufficiently larger than the electroweak scale ($v$) but still much smaller than the scale of the new physics ($\Lambda$), $v\ll E\ll \Lambda$.  
Focusing on the dimension-6 operators, we find the sum rules to provide useful insight on the connection between low energy observables and the properties of new physics beyond the SM, and are consistent with the usual matching and running procedure of EFT for the cases we study.  Interestingly, we also find it possible to impose symmetries at the amplitude level to suppress the contribution of certain dimension-6 operators, which can be connected to the familiar custodial symmetries of the SM Higgs and fermion sectors.

The rest of this paper is organized as follows:  In \autoref{sec:osamp} we briefly review the sum rules before providing a classification of the forward elastic amplitudes in the SMEFT.   
We then perform a systematic enumeration of the sum rules for the dimension-6 operators in \autoref{sec:sum} and discuss their implications in \autoref{sec:imp}.  Finally, we apply the sum rules to a few benchmark models in \autoref{sec:bm}. Our conclusion is drawn in \autoref{sec:con}. 
A short review on the essential results of the spinor helicity formalism is provided in \autoref{app:heli}, and a derivation of the forward limit is given in \autoref{sec:afor}.

%%%%%%%%%%%%%%%%%%%%%%%%%%%%%%%%%%%%%%%%%%%%%%%%%%%%%%%%%%%%%%%%%%
\section{Sum rules and elastic amplitudes}
\label{sec:osamp}
%%%%%%%%%%%%%%%%%%%%%%%%%%%%%%%%%%%%%%%%%%%%%%%%%%%%%%%%%%%%%%%%%%
\subsection{Sum rules}
%%%%%%%%%%%%

The starting point for writing down a sum rule is to 
consider the elastic scattering of two particles (denoted as $a$ and $b$) and write down the amplitude in the forward limit, 
which is a function of the Mandelstam variable $s$ alone due to the relation~$s+t+u =4m^2$,\footnote{  We  assume $a$ and $b$ have the same mass $m$ for convenience. }
\begin{equation}
\tilde{\A}_{ab}(s) \equiv \left. \A (ab \, \to \, ab) \right|_{t=0} \,. \label{eq:A00}
\end{equation}
Throughout our paper we will use $\tilde{\A}$ to denote amplitudes in the forward limit to distinguish them from the general amplitudes $\A$. Performing an analytical continuation of $s$ to the complex plane, and expanding $\tilde{\A}_{ab}$ around the point $s=\mu^2$, we obtain
\begin{equation}
\tilde{\A}_{ab} (s) = \sum_n c_n (s-\mu^2)^n \,, \hspace{1.5cm}% ~~~~~\mbox{where}~~ 
c_n  = \frac{1}{2\pi i} \oint\limits_{s=\mu^2} ds \frac{\tilde{\A}_{ab} (s)}{(s-\mu^2)^{n+1}}  \,,  \label{eq:A02}
\end{equation}
where each coefficient $c_n$ is written as a contour integral around the point $s = \mu^2$.  Expanding the contour to infinity, 
one picks up all the non-analytic structures in the complex $s$-plane.  For an interacting theory, discontinuities on the real axis generally exist, which can be related to the total cross sections of the scattering of $a$ and $b$ (and $\bar{b}$, the anti-particle of $b$) via the optical theorem. This gives a dispersion relation in the following form\footnote{See {\it e.g.} Refs.~\cite{Bellazzini:2016xrt, Bellazzini:2014waa} for a more detailed derivation, in particular on how the crossing symmetry leads to the term $\sigma^{a\bar{b}}_{\rm tot}$.  Note also that in \autoref{eq:A03} we have omitted possible additional IR poles from SM contributions, which are discussed later in \autoref{sec:valid}. } 
\begin{equation}
c_n   = \int^\infty_{4m^2} \frac{ds}{\pi}  s\sqrt{1-\frac{4m^2}{s}} \left( \frac{\sigma_{\rm tot}^{ab} }{ (s-\mu^2)^{n+1} } + (-1)^n \frac{\sigma_{\rm tot}^{a\bar{b}}}{(s-4m^2+\mu^2)^{n+1}}  \right) + c^\infty_n \,, \label{eq:A03}
\end{equation}
where $\sigma^{ab}_{\rm tot}$ ($\sigma^{a\bar{b}}_{\rm tot}$) is the total cross section of the scattering of particles $a$ and $b$ ($\bar{b}$).  The $\sigma^{a\bar{b}}_{\rm tot}$ term corresponds to contributions in the $u$-channel, which is mapped to the $s$-plane and rewritten using the $s\leftrightarrow u$ crossing relation which exchanges $b$ and $\bar{b}$. 
The contour at infinity gives the term $c^\infty_n \equiv  \frac{1}{2\pi i} \oint\limits_{s\to \infty} ds \frac{\A (s)}{(s-\mu^2)^{n+1}}$, which vanishes for $n>1$  as a result of the Froissart bound~\cite{Froissart:1961ux}.  The $c_n$s are closely related to the Wilson coefficients in the EFT.  As we will show later, the expansion in \autoref{eq:A02} exactly maps to the EFT expansion in the $m^2 \ll \mu^2 \ll \Lambda^2$ limit, with $n=1,\, 2,\, ....$ corresponding to operator dimensions 6, 8, ...., respectively.  This limit is consistent with our massless SMEFT assumption. The scale $\mu$ can be considered as the energy at which the relevant parameters in the scattering are defined.  For an even $n$, the two cross section terms in \autoref{eq:A03} are both positive, implying that $c_n$ must be positive for a nontrivial theory.  For $n=1$ which corresponds to the dimension-6 operators,  the two cross section terms have opposite signs, and the boundary term $c^\infty_1$ can also be nonzero in general.

%%%%%%%%%%%%%%%%%%%%%%%%%%%%%%%%%%%%%%%%%%%
\subsection{Elastic helicity amplitudes in the SMEFT}
\label{sec:eaeft}
%%%%%%%%%%%%%%%%%%%%%%%%%%%%%%%%%%%%%%%%%%%

The Standard Model Effective Field Theory (SMEFT) is obtained by augmenting the SM Lagrangian with higher dimensional operators comprised of only the SM field content, which form an expansion in terms of the inverse of some energy scale $\Lambda$.  Assuming baryon and lepton numbers are conserved, only operators of even dimensions are allowed, and the SMEFT Lagrangian can be written as\footnote{ The only dimension-5 operator in the SM is the Weinberg operator of the form $LLHH$.  In the massless limit it contributes to neither 3-point amplitudes nor 4-point elastic amplitudes.} 

\begin{equation}
\mathcal{L}_{\rm SMEFT} =
    \mathcal{L}_{\rm SM} + 
    \sum_i \frac{c^{(6)}_i}{\Lambda^2} \mathcal{O}^{(6)}_i + 
    \sum_j \frac{c^{(8)}_j}{\Lambda^4} \mathcal{O}^{(8)}_j +         \cdots \,.   \label{eq:smeft}
\end{equation}
We consider how these higher dimensional operators could contribute to an elastic scattering amplitude in the massless limit.  
The dimension of an amplitude is given by $[\A_n] = 4-n$, where $n$ is the number of external legs.  Any $n$-point amplitude can receive contributions from different couplings in the theory. Ordering the contributions by the dimension of the couplings, $\A_n$ can be expanded as
\begin{equation}
\A_n = \sum_i g_{[i]} \A^{[4-n-i]}_n \,,   
\end{equation}
where $i$ denotes the dimension of the coupling $g_{[i]}$.  $\A^{[4-n-i]}_n$ is the contribution to the amplitude proportional to $g_{[i]}$, and must have dimension $4-n-i$ as denoted in the superscript, so that the total dimension equals $4-n$.   
A dimension-$d$ operator in the Lagrangian has a coupling with dimension $i=4-d$. Hence, it can contribute to the amplitude $\A_n$ a term in the form of  $g_{[4-d]} \A^{[d-n]}_n$.\footnote{This mapping can be spoiled if the fields develop vacuum expectation values (vevs), as in the SM. Working in the massless limit, we do not consider the Higgs vev here.}  We also note that in the massless limit, all couplings in the SM are dimensionless.   A 4-point amplitude can thus be written as
\begin{equation}
\A_4 =  g_{[0]} \A^{[0]}_4 +   g_{[-2]} \A^{[2]}_4 + g_{[-4]} \A^{[4]}_4 +....  \label{eq:ampexp}
\end{equation}
where $\A^{[0]}_4$ is the SM contribution, $\A^{[2]}_4$ is obtained with one insertion of dimension-6 operators (with $g_{[-2]}\propto 1/\Lambda^2$), and $\A^{[4]}_4$ comes from either one insertion of dimension-8 operators or two insertions of dimension-6 operators.  Similarly, a 3-point amplitude can be written as
\begin{equation}
 \A_3 =  g_{[0]} \A^{[1]}_3 +   g_{[-2]} \A^{[3]}_3 + ... \label{eq:amp3exp}
\end{equation}
Let us look at the $\A^{[2]}_4$ term of the 4-point amplitude.  The claim is that the only kind of contribution to the elastic scattering is in the form of a 4-point contact interaction.\footnote{Strictly speaking, this only applies in the case of tree level EFT contributions to the amplitude. Loop contributions can give rise to amplitudes with different momentum structures. } 
In this case, locality of the EFT dictates that there is no momentum dependence on the denominator.  This is obviously the case when the amplitude is built from one dimension-6 operator with 4 external particles. 
 
\begin{figure}[t]
\centering
\includegraphics[width=0.32\textwidth]{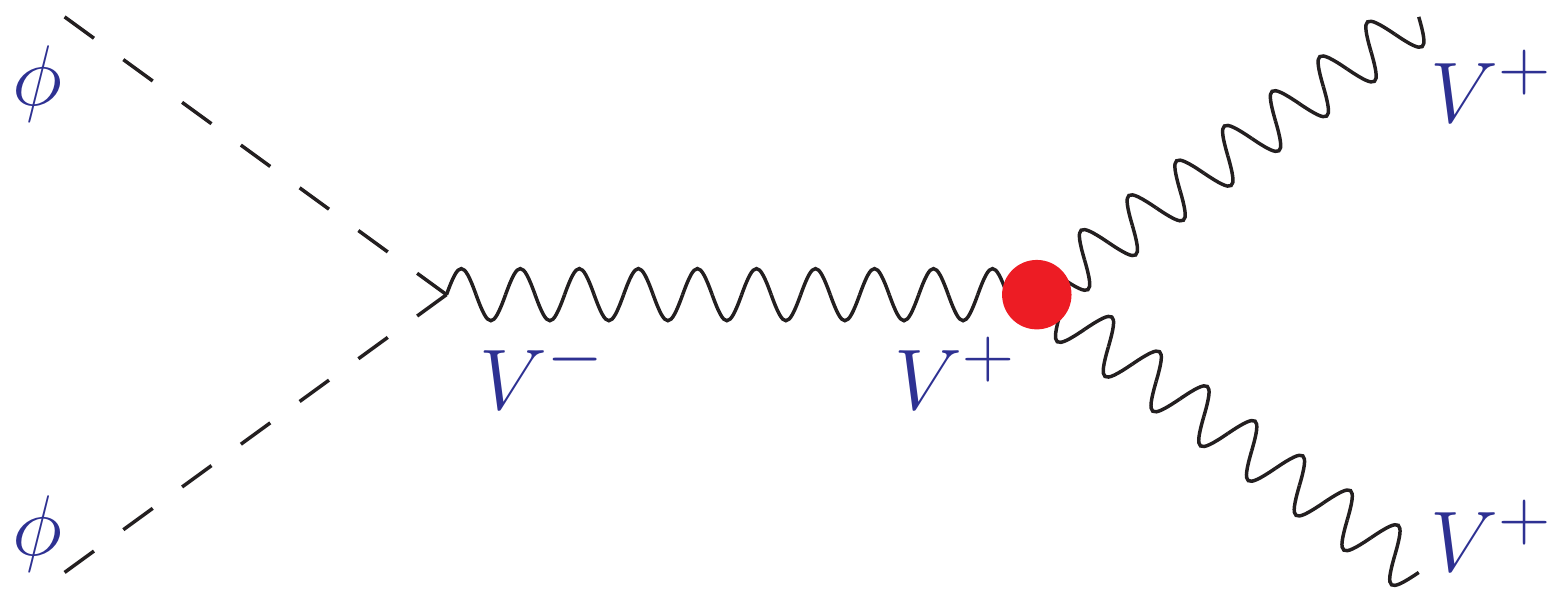} \hspace{0.05cm}
\includegraphics[width=0.32\textwidth]{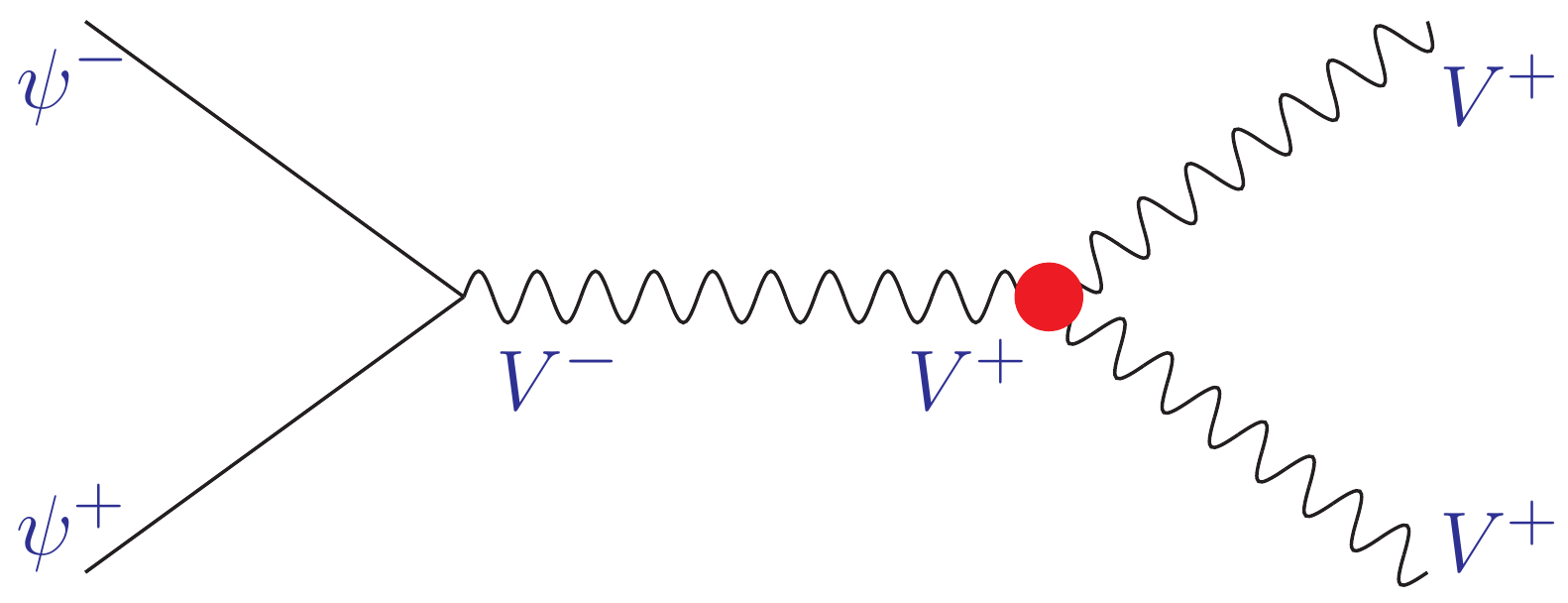}  \hspace{0.05cm}
\includegraphics[width=0.32\textwidth]{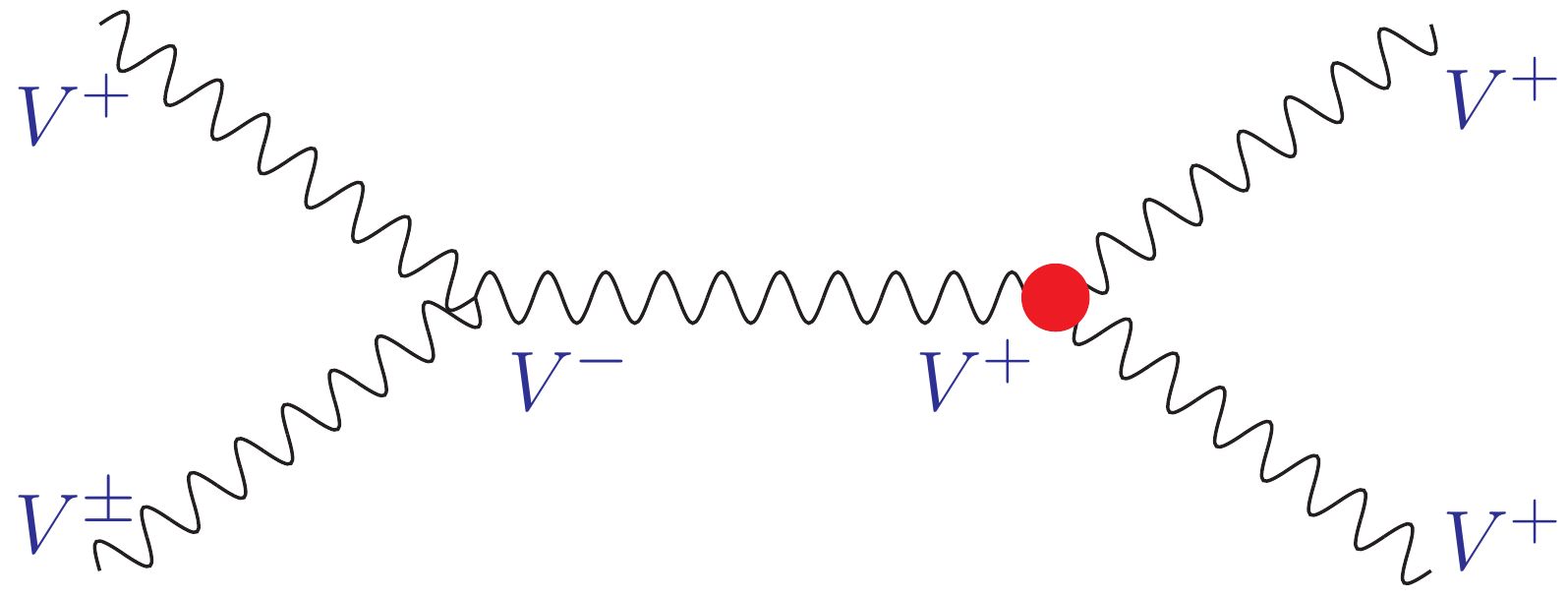}
\caption{Possible factorization channels of $\A^{[2]}_4$ that contain a $\A(V^+V^+V^+)$ part.  The superscript of a particle indicates the sign of its helicity, with the all-in/all-out convention.  The diagrams with $\A(V^-V^-V^-)$ can be obtained by flipping all the helicities.  The red dot denotes an insertion of a dimension-6 operator.  None of the processes are elastic in the helicity basis.  Note in particular that $\A(V^-V^-V^+V^+)$ can not be generated by only one insertion of dimension-6 operators.    }
\label{fig:a4vvv}
\end{figure}

 A 4-point amplitude can also be built by combining two 3-point amplitudes.  It can only factorize into the form $\sim\A^{[1]}_3 \, \frac{1}{p^2} \, \A^{[3]}_3$ in the SMEFT, where $\A^{[1]}_3$ is from SM and $\A^{[3]}_3$ is generated by one insertion of dimension-6 operators, as shown in \autoref{eq:amp3exp}.  Factorizations of the form  $\sim\A^{[2]}_3 \, \frac{1}{p^2} \, \A^{[2]}_3$, corresponding to two insertions of d5 operators, are not possible within the SMEFT. Assuming all particles have spin less than or equal to one ($|h|\leq1$), the only $\A^{[3]}_3$ we can write down is from three vectors with the same helicity, $\A(V^+V^+V^+)$  or $\A(V^-V^-V^-)$.\footnote{See \autoref{app:heli} for a short derivation of this statement.  Here the superscript of a particle indicates the signs of its helicity.  We also use the convention that all particles are going in the vertex (or all are going out).} 
They are generated by the operators
\begin{equation}
\mathcal{O}_{3W}= \frac{1}{3!} g \epsilon_{abc} W^{a\,\nu}_\mu W^b_{\nu \rho} W^{c\,\rho\mu} \,,  ~~~~~~ \mbox{and} ~~~~~~ \mathcal{O}_{3\widetilde{W}}= \frac{1}{3!} g \epsilon_{abc} \widetilde{W}^{a\,\nu}_\mu W^b_{\nu \rho} W^{c\,\rho\mu}\,,  \label{eq:o3w}
\end{equation}
for the electroweak gauge bosons, 
or $\mathcal{O}_{3G}$ and $\mathcal{O}_{3\widetilde{G}}$ for the gluons.
 One could then attach a pair of scalars ($\phi$), fermions ($\psi$) or vectors ($V$) on one of the vector to make a 4-point amplitude, as shown in \autoref{fig:a4vvv}.  
 However, note that none of these amplitudes is elastic.  This is because we need the incoming and outgoing particle to have the same helicity, {\it e.g.} $V^+ \to V^+$.  In the all-in/all-out convention, the same particle thus must have opposite helicities.  This is not the case for any of the processes in \autoref{fig:a4vvv}.  We thus conclude that an elastic $\A^{[2]}_4$ can only be of the contact form.

Our conclusion is seemingly in contradiction with the fact that other operators can also generate 3-point interactions.  In particular, operators
\begin{align}
\mathcal{O}_{HW} =&~  ig(D^\mu H)^\dagger \sigma^a (D^\nu H) W^a_{\mu\nu} \,, \hspace{1cm}
\mathcal{O}_{HB} =  ig'(D^\mu H)^\dagger  (D^\nu H) B_{\mu\nu} \,, \nonumber \\
\mathcal{O}_{W} =&~  \frac{ig}{2}(H^\dagger \sigma^a \overleftrightarrow{D_\mu} H) D^\nu W^a_{\mu\nu} \,,  \hspace{1.5cm}
\mathcal{O}_{B} =  \frac{ig'}{2}(H^\dagger \overleftrightarrow{D_\mu} H) \partial^\nu B_{\mu\nu} \,,  \label{eq:ohw}
\end{align}
would generate $\phi\phi V$-type couplings.  Equivalently, they contribute to the anomalous triple gauge couplings once the Higgs boson develop a vev.  However, note that in the massless limit they do not generate on-shell 3-point amplitudes, since these $\phi\phi V$ couplings have a $p^2$ dependence which vanishes on shell.  While they still contribute to 4-point amplitudes via an off-shell 3-point amplitude, the $p^2$ in the propagator of  $\A^{[1]}_3 \, \frac{1}{p^2} \, \A^{[3]}_3$ is cancelled by the $p^2$ from $\A^{[3]}_3$, so the 4-point amplitude generated in this way is still a contact one ({\it i.e.} it does not have physical factorization channels).  It is not a coincidence that these operators can be exchanged with operators with more vertices by applying integration by parts and the equations of motion of the Gauge bosons, as done in the Warsaw basis \cite{Grzadkowski:2010es}.

\begin{table}[t]
\centering
\begin{tabular}{|c|c|c|} \hline
elastic 4-point amplitudes & spinor form of $\A^{[2]}_4$ & spinor form of $\A^{[4]}_4$   \\  
$\A(1\,2 ~\to~ 3_{=1} \, 4_{=2} )$   &  (d6 operators)  &  (d8 or d6$^2$)    \\ \hline\hline
$\phi_1\phi_2 \phi^*_1 \phi^*_2$   &   $s_{ij}$  &  $s_{ij}\times s_{kl}$    \\ \hline
$\psi^- \phi \psi^+ \phi^*$ & $\la 1 2\ra[23]$  &   $\la 1 2\ra[23] \times s_{ij} $   \\ \hline
$\psi^-_1 \psi^-_2 \psi^+_1 \psi^+_2$ & $\la 1 2\ra[34]$  &  $\la 1 2\ra[34] \times s_{ij} $  \\ \hline
$V^- \phi V^+ \phi^*$ & \xmark   &  $\la 1 2\ra^2 [23]^2$   \\ \hline
$V^- \psi^- V^+ \psi^+$ & \xmark  &  $\la 1 2 \ra^2 [23][34]$    \\ \hline
$V^-_1 V^-_2 V^+_1 V^+_2$ & \xmark   &     \,$\la 1 2\ra^2 [34]^2\,,~\la 1 2\ra^2 [34]^2\frac{t-u}{s}$ \\ \hline 
\end{tabular}
\caption{ A full list of all possible 4-point elastic scattering amplitudes in the helicity basis, with the corresponding tree-level form with mass dimensions 2 (one insertion of dimension-6 operators) and 4 (dimension-8 or dimension-6-squared) in the massless case.  We use $\phi$, $\psi$ and $V$ to denote scalars, fermions and vectors, respectively.  The $+/-$ signs denote helicities in the usual all-in/all-out convention.  The ordering of the particles is such that the incoming particle 1 is outgoing particle 3, and 2 is 4.  The labels 1 and 2 are explicitly shown in the subscripts for scattering of the same particle type.   The forward limit corresponds to $t=s_{13} \to 0$.  
Amplitudes that can be obtained by crossing (such as $\A(\psi^-_1\psi^+_2\psi^+_1\psi^-_2)$) are not explicitly shown. 
The \xmark~mark denotes that one can not write down a term that fulfills all the consistency requirements. 
$s_{ij}$ denotes a general linear function of the Mandelstam variables in the form $c_s s + c_t t + c_u u$, where $c_{s,t,u}$ are constants. }
\label{tab:a4elastic}
\end{table}

Since the 4-point amplitudes must be contact, their kinematic forms are strongly constrained by a number of requirements.  
Namely, all the angle and square brackets have to be in the numerator, the total dimension of $\A^{[2]}_4$ should be 2, and the little group scaling needs to be consistent with the helicities of the particles. 
We list in \autoref{tab:a4elastic} all the possible 4-point elastic amplitudes, with their spinor form for $\A^{[2]}_4$ up to some couplings constants.  Amplitudes that can be obtained by the crossing $s\leftrightarrow u$ (exchanging particles $1\leftrightarrow 3$ or $2\leftrightarrow 4$) are not explicitly shown.  
We note again that, in the all-in/all-out convention,  the elasticity of  a massless amplitude enforces it to have zero net helicity.   In other words, all the amplitudes in \autoref{tab:a4elastic} must have equal numbers of square and angle brackets.

The spinor forms of $\A^{[2]}_4$ in \autoref{tab:a4elastic} have some remarkable features.  First, with the exception of the 4-scalar amplitude, they are completely fixed by little group scaling.\footnote{Note that certain combinations of spinor products can be related to each other and are not independent.  For instance, momentum conservation imposes $\la 1 2\ra[23] = - \la 1 4\ra[43]$, as shown in \autoref{eq:a1223}. }  In particular, for $\A(V^- \phi V^+ \phi^*)$, $\A(V^- \psi^- V^+ \psi^+)$ and $\A(V^-_1 V^-_2 V^+_1 V^+_2)$ (and the ones related by crossing), we simply could not write down an $\A^{[2]}_4$ term that fulfills all the consistency requirements.
This means that dimension-6 operators could not contribute to these amplitudes at tree level.\footnote{ Even at the one-loop level, the dimension-6 operators could only have rational contributions to these amplitudes as a result of the helicity selection rules~\cite{Craig:2019wmo}. }   For the 4-scalar amplitude, we note that $\A^{[2]}_4$ can only be linear in terms of the Mandelstam variables $s$, $t$ and $u$, since it has a mass-dimension two and is also invariant under little group scaling.  The massless relation $s+t+u=0$ is not sufficient to fix $\A^{[2]}_4$.

It is straightforward to repeat the analysis for the $\A^{[4]}_4$ terms, which we also list in \autoref{tab:a4elastic}. They correspond to one insertion of dimension-8 operators or two insertions of dimension-6 operators.  
An important observation is that a 3-point massless on-shell amplitude could not be generated by operators of dimension higher than 6, assuming all particles have spins less than or equal to one.   
We thus arrive at the similar conclusion that $\A^{[4]}_4$ terms have to come from contact 4-point interactions generated by dimension-8 operators, with the exception that $\A(V^-_1 V^-_2 V^+_1 V^+_2)$ can now be generated with two insertions of dimension-6 operators, by combining $\A_3(V^-V^-V^-)$ and $\A_3(V^+V^+V^+)$.   
This generates a pole in the $s$-channel, while requiring the amplitude to be antisymmetric under $1\leftrightarrow 2$ or $3\leftrightarrow 4$ (as the 3-point amplitudes are anti-symmetric) gives the spinor form shown in \autoref{tab:a4elastic}.
Similarly, we could also conclude that the higher order terms in the amplitude expansion ($\A^{[6]}_4$, $\A^{[8]}_4$, ...) must come from contact 4-point interactions.

A potential caveat of the helicity-amplitude approach is that it does not exhaust all possible elastic amplitudes.  
While physics is obviously independent of the basis for particle states, the notion of elasticity is not.  For instance, a 4-vector amplitude with different initial and final state helicities ({\it e.g.} $V^+V^+ \to V^-V^-$)  is inelastic in the helicity basis.  By changing to the linear basis, it would contribute to elastic amplitudes.  
It is shown explicitly in Ref.~\cite{Remmen:2019cyz} that for the 4-vector amplitudes, certain positivity bound involving CP-odd dimension-8 operators can be written down in the linear basis but is absent in the helicity basis.  
The transversity basis in Ref.~\cite{deRham:2017zjm} is also useful for amplitudes of spinning particles. 
Equivalently, this requires one to also consider inelastic amplitudes with vectors in the helicity basis, in which case the interpretation of the dispersion relations is much less straightforward.  However, we note that these inelastic amplitudes do not contribute to nontrivial sum rules at the $\A^{[2]}_4$ level.
For both contact $V \phi V \phi$ and $V \psi V \psi$ amplitudes, one could show that with two $V^+$s (or two $V^-$s), the amplitude must vanish in the forward limit due to angular momentum conservation. 
Similarly, 4-vector amplitudes with an odd number of $V^+$ or $V^-$ ({\it e.g.} $\A(V^-V^+V^+V^+)$) must also vanish in the forward limit.  %
It seems possible to write down a nonzero forward amplitude for $\A(V^+V^+V^+V^+)$ and $\A(V^-V^-V^-V^-)$.  As we will show later, a massless $\A^{[2]}_4$ amplitude is an odd function of $s$ in the forward limit and vanishes if it is symmetric under the crossing $s \leftrightarrow u$.  This is the case with linear polarizations, if $V$ is its own anti-particle ({\it e.g.} $W^0$ or gluon).  As such, one is left with the scattering of $W^+ W^- \to W^+ W^-$ (or similar combinations of gluons) with linear polarizations, which could generate a sum rule for the operators $\mathcal{O}_{3W}$ and $\mathcal{O}_{3\widetilde{W}}$.  
However, a crucial observation is that the one-loop contributions to $c_{3W}$ and $c_{3G}$ have opposite signs for boson loops and fermion loops with the same group representation~\cite{Henning:2014wua}.  Without further investigations (which we leave to future studies),  we could already confirm the non-existence of a consistent sum rule for these operators from elastic amplitudes, since the cross section terms on the righthand side of the sum rule \autoref{eq:A03} cannot generate such an opposite sign between fermion and boson final states. 
For the sum rules to provide useful information on the properties of the heavy particles in the full theory, such as their charges, we restrict ourselves to the scattering of states with definitive Poincar\'e representations and quantum numbers ({\it i.e.}~the usual notion of particles).  
With this restriction, the SM fermions in the unbroken electroweak phase ($q_L, u_R, d_R, \ell_L, e_R$) do not mix with each other since they all have different quantum numbers (assuming one generation), and the helicity basis is sufficient to describe them.    
As such, we conclude that the helicity basis is sufficient for the enumeration of all sum rules of dimension-6 operators.  

We will focus on the forward limit of the elastic scattering amplitudes in deriving the sum rules. 
It can be shown that, for massless particles with any spins, 
the forward elastic amplitudes in the helicity basis 
are always invariant under the little group scaling and can be treated as if they are scalar amplitudes~\cite{Bellazzini:2016xrt}.  A short derivation for this result is presented in \autoref{sec:afor}. 
For the terms in \autoref{tab:a4elastic}, we then have
\begin{equation}
\tilde{\A}^{[2]}_4 \equiv \A^{[2]}_4 |_{t\to0} \propto s  \,, \hspace{1cm}  \tilde{\A}^{[4]}_4 \equiv  \A^{[4]}_4 |_{t\to0} \propto s^2 \,.  \label{eq:foamps}
\end{equation}
Comparing \autoref{eq:foamps} with \autoref{eq:A02} and \autoref{eq:A03}, we note that 
$\tilde{\A}^{[2]}_4$ and $\tilde{\A}^{[4]}_4$ match the $n=1$ and $n=2$ cases of the sum rule in \autoref{eq:A03} in the limit $\mu \to 0$.  Since we work in the massless limit, the mass term $m$ in \autoref{eq:A03} should also be set to zero, which may lead to issues such as analyticity at the point $s=0$ and potential IR divergences of loop corrections.
However, it should be understood that we are working in the limit $m^2 \ll \mu^2 \ll \Lambda^2$ (with $\mu^2$ having a small non-zero imaginary part to avoid the branch cut on the real axis), rather than the exact massless case $m=\mu=0$.  The leading order contribution of finite $m$ comes from the SM, and is suppressed by powers of $m^2/\mu^2$, while the contributions from higher dimensional operators are further suppressed by powers of $\mu^2/\Lambda^2$. These contributions are omitted in our study.

%==============================

%%%%%%%%%%%%%%%%%%%%%%%%%%%%%%%%%%%%%%%%%%%%%%%%%%%%%

%%%%%%%%%%%%%%%%%%%%%%%%%%%%%%%%%%%%%%%%%%%%%%%%%%%%%%%%%%%%%%%%%%
\section{Sum rules of dimension-6 operators}
\label{sec:sum}
%%%%%%%%%%%%%%%%%%%%%%%%%%%%%%%%%%%%%%%%%%%%%%%%%%%%%%%%%%%%%%%%%%

Having established the connection between the helicity amplitudes and sum rules, we are now ready to write down the sum rules in the SMEFT.  
We will be focusing on the ones relevant for the dimension-6 operators.  
As mentioned in the previous section, they correspond to the $n=1$ term in \autoref{eq:A03} in the limit $m^2 \ll \mu^2 \ll \Lambda^2$,  
\begin{equation}
 \left. \frac{d \tilde{\A}_{ab}(s)}{d s}  \right|_{s=0} =  \int^\infty_{0}  \frac{ds}{\pi s} \left( \sigma^{ab}_{\rm tot} - \sigma^{a\bar{b}}_{\rm tot}  \right)  + c_\infty \,. \label{eq:sum1}
\end{equation}
We can now replace particles $a$ and $b$ with the SM particles.  As suggested by \autoref{tab:a4elastic}, one only needs to consider the scattering between two Higgs, two fermions, or one Higgs and one fermion. 
In each case, it is important to establish the connection between the number of independent parameters in the theory and the number of independent sum rules they are subject to.  As such, we will perform the counting directly based on amplitudes, and only make connections to the operator coefficients afterwards.  
A general principle of this approach is that the number of independent parameters for a particular amplitude is given by the number of independent kinematic form it can have, which is dedicated by the little group scaling as well as symmetries of the amplitude~\cite{Shadmi:2018xan, Ma:2019gtx}.  As an illustration, let us look at a few examples for the 4-point scalar amplitude.  
As shown in \autoref{tab:a4elastic}, its $\A^{[2]}_4$ term 
is a linear combination of the Mandelstam variables, and can be written in the general form $c_s \, s + c_t \, t + c_u \, u$.  Consider first the case of a single real scalar, the amplitude should be symmetric under any exchange between $s$, $t$ or $u$, {\it i.e.} $c_s=c_t=c_u$.  Combined with the massless relation  $s+t+u=0$, we could conclude that $\A^{[2]}_4$ must vanish.  This is consistent with the fact that the dimension-6 operator for the single real scalar that contributes to the $\A^{[2]}_4$ term is redundant and can be eliminated by a field redefinition~\cite{Shadmi:2018xan}.  Similarly, for a single complex scalar, symmetry requires that $\A(\phi\phi \phi^* \phi^*)$ is invariant under the exchange of the two $\phi$s (or $ \phi^*$s).  Therefore, $\A(\phi\phi \phi^* \phi^*)$ is symmetric under the exchange $t\leftrightarrow u$ and can only be proportional to $s$.  Not surprisingly, there is also only one independent dimension-6 operator for the single complex scalar.

Starting with the $\A^{[2]}_4$ amplitudes in \autoref{tab:a4elastic}, we then go through the following procedure to count the sum rules:
\begin{itemize}

   \item Count the number of independent amplitudes.  SM particles fill various gauge multiplets, and fermions can also come in with different flavors.  One needs to properly count the degrees of freedoms in order to derive all the sum rules. 
   
   \item  For each independent amplitude, count the number of independent parameters in it.   If the amplitude contains one parameter, and does not vanish in the forward limit, it then produces one sum rule for this parameter.

   \item If an amplitude contains more than one independent parameter, it is possible to also obtain multiple sum rules by considering different physical states.  In particular, one could symmetrize the amplitude with respect to the Mandelstam variables, and obtain different forward limits.
   The Higgs-Higgs amplitude below provides an explicit example of this.  
   
\end{itemize}     
For simplicity, we only include one generation of fermions, while the generalization to 3 generations is straightforward but somewhat tedious.  A recent study on the flavor constraints from dimension-8 four-fermion operators can be found in Ref.~\cite{Remmen:2020vts}.  
Given that we are working in the massless limit with no EWSB, we will parameterize the Higgs doublet as $H= \bpm \phi^+ \\ \phi^0 \epm$ and work directly with the complex components $\phi^+$ and  $\phi^0$, together with $H^{\dagger}= \bpm \phi^- & \phi^{0*} \epm$ where $ \phi^- = (\phi^+)^*$.  
Following the procedure above, we list the sum rules below for each of the three types of amplitudes.

%%%%%%%%%%%%%%%%%%%%%%
\subsection{Higgs-Higgs amplitudes}

The only scalar in the SM is the Higgs doublet.  Writing down the 4-Higgs amplitude (with the all-in/all-out convention) with explicit $SU(2)$ indices, $\A(H_i H_j H^\dagger_k H^\dagger_l)$, 
gauge invariance then requires that we contract the $SU(2)$ indices, with either $i=k,\, j=l$ or $i=l,\, j=k$.  We immediately realize that the two amplitudes produced by these two contractions are not independent, but are related by an exchange of the two $H$s (or the two $H^\dagger$s).  We thus have only one independent amplitude.  Letting $i=k \neq j=l$, we can write its $\A^{[2]}_4$ term with two independent paraemeters,
\begin{equation}
\A(H_i H_j H^\dagger_i H^\dagger_j) = c_s \, s + c_u \, u  \,,  \label{eq:a4h6a}
\end{equation}
where the $t$ term is eliminated via the relation $s+t+u=0$.  \autoref{eq:a4h6a} is an elastic amplitude, and gives a sum rule on $c_s - c_u$ in the forward limit $t=0$, $u=-s$. 
However, a different forward limit can be obtained by taking $i=k = j=l$ which symmetrizes \autoref{eq:a4h6a} under $t \leftrightarrow u$, giving the elastic amplitude
\begin{equation}
\A(H_i H_i H^\dagger_i H^\dagger_i) = 2 c_s \, s + c_u \, t + c_u \,u  = (2 c_s - c_u) s  \,,  \label{eq:a4h6b}
\end{equation}
which instead gives a sum rule on the combination $2 c_s -c_u$.  Another possibility is to let $i=k = j=l$ and only take the real component, making the amplitude totally symmetric under $s$, $t$ and $u$.  However, the $\A^{[2]}_4$ term vanishes in this case,  as discussed in the single real scalar case above.  In fact, no additional independent sum rule can be written down in this case.  \autoref{eq:a4h6a} and \autoref{eq:a4h6b} thus contain two independent parameters and gives two sum rules on the combinations,
\begin{equation}
c_s - c_u \, ,  \hspace{1cm}  2c_s - c_u \,.  \label{eq:hsum01}
\end{equation}

To connect \autoref{eq:hsum01} with the dimension-6 operators, one simply needs to compute the amplitudes \autoref{eq:a4h6a} and \autoref{eq:a4h6b} in a given operator basis.  Only two independent dimension-6 operators contribute to the $\A^{[2]}_4$ term of the 4-Higgs amplitude, which can be chosen as $\mathcal{O}_H$ and $\mathcal{O}_T$ in \autoref{tab:op1}.  
With an explicit calculation, we obtain
\begin{align}
\A^{[2]} (\phi^+ \phi^- \to \phi^+ \phi^- )  =&~ \frac{c_H+3c_T}{\Lambda^2} s \,,  \label{eq:a4hO1} \\
\A^{[2]} (\phi^+ \phi^0 \to  \phi^+ \phi^0)  =&~ -\frac{c_H+c_T}{\Lambda^2}s - \frac{c_H-c_T}{\Lambda^2} u  \,,  \label{eq:a4hO2}  
\end{align}
where, with an abuse of notation, we have absorbed the couplings (with mass dimension $-2$) in the amplitudes.  It is then straightforward to make the connection
\begin{equation}  
c_s \to   -\frac{c_H+c_T}{\Lambda^2}  \,, \hspace{1cm}  c_u \to - \frac{c_H-c_T}{\Lambda^2} \,.
\end{equation}
The two sum rules are given by
\begin{align}
\frac{c_H + 3c_T}{\Lambda^2} =  \left. \frac{d \tilde{\A}_{\phi^+ \phi^-}}{d s} \right|_{s=0}  =&~ \int^\infty_{0} \frac{ds}{\pi  s} \left( \sigma^{\phi^+ \phi^-}_{\rm tot} - \sigma^{\phi^+ \phi^+}_{\rm tot}   \right) + c_\infty \,, \label{eq:sumhh1} \\
- \frac{2 c_T}{\Lambda^2} =  \left. \frac{d \tilde{\A}_{\phi^+ \phi^0}}{d s} \right|_{s=0} =&~   \int^\infty_{0}  \frac{ds}{\pi s} \left( \sigma^{\phi^+ \phi^0}_{\rm tot} - \sigma^{\phi^+ \phi^{0*}}_{\rm tot}  \right)  + c_\infty \,, \label{eq:sumhh2}
\end{align}
where in the second equation the $c_H$ terms are cancelled in the forward limit.

\begin{table}[t]
\centering
\begin{tabular}{l|l} \hline \hline
&\\[-0.4cm]
$\mathcal{O}_H = \frac{1}{2} (\partial_\mu |H|^2)^2 $  & $\mathcal{O}_T =  \frac{1}{2} (H^\dagger \overleftrightarrow{D_\mu} H)^2$   \\ &\\[-0.4cm]  \hline%\hline
&\\[-0.4cm]
 $\mathcal{O}_{H\ell} = i  H^\dagger \overleftrightarrow{D_\mu} H \bar{\ell}_L \gamma^\mu \ell_L  $  &    \\
 $\mathcal{O}'_{H\ell} = i H^\dagger \sigma^a \overleftrightarrow{D_\mu} H \bar{\ell}_L \sigma^a \gamma^\mu \ell_L $  &   $\mathcal{O}_{He} = i H^\dagger \overleftrightarrow{D_\mu} H \bar{e}_R \gamma^\mu e_R $  \\ \hline%\hline
&\\[-0.4cm]
$\mathcal{O}_{Hq} = i  H^\dagger \overleftrightarrow{D_\mu} H \bar{q}_L \gamma^\mu q_L  $ &  $\mathcal{O}_{Hu} = i H^\dagger \overleftrightarrow{D_\mu} H \bar{u}_R \gamma^\mu u_R $ \\
$\mathcal{O}'_{Hq} = i H^\dagger \sigma^a \overleftrightarrow{D_\mu} H \bar{q}_L \sigma^a \gamma^\mu q_L $ &  $\mathcal{O}_{Hd} = i H^\dagger \overleftrightarrow{D_\mu} H \bar{d}_R \gamma^\mu d_R $ \\  \hline \hline
\end{tabular}
\caption{The dimension-6 operators related to Higgs-Higgs and Higgs-fermion sum rules.}
\label{tab:op1}
\end{table}

%%%%%%%%%%%%%%%%%%%%%%
\subsection{Higgs-fermion amplitudes}

Again, we proceed by writing down the amplitudes with all possible ways to contract the group indices.  If the fermion $f$ is an $SU(2)$ singlet ($f=\, u_R, \, d_R, \,e_R$), the $SU(2)$ indices can only be contracted between the two Higgs, with only one independent amplitude, $\A(H_i f H^\dagger_i \bar{f})$.  If $f$ is an $SU(2)$ doublet ($f=\,q_L,\,l_L$), one could contract the $SU(2)$ indices in two ways, giving $\A(H_i f_j H^\dagger_i \bar{f}_j)$ and $\A(H_i f_j H^\dagger_j \bar{f}_i)$.  The latter is not elastic if $i\neq j$, while the elastic amplitude $\A(H_i f_i H^\dagger_i \bar{f}_i)$ receives contribution from both contractions. Therefore, we have two independent elastic amplitudes which are $\A(H_i f_j H^\dagger_i \bar{f}_j)$ $(i \neq j)$ and  $\A(H_i f_i H^\dagger_i \bar{f}_i)$.    
Having $SU(3)$ indices does not change the counting, since they can only contract between the two quarks.  We note from \autoref{tab:a4elastic} that the kinematic structure of $\A^{[2]}_4$ is fixed for the scalar fermion amplitudes, so each independent amplitude has one parameter in the $\A^{[2]}_4$ term, which is subject to one sum rule.  This gives a total number of $2\times 2 +3 =7$ sum rules for each family of SM fermions.  Not surprisingly, they can be connected to the  
7 $\mathcal{O}_{Hf}$ type operators in \autoref{tab:op1}.  The 4 sum rules for the quarks can be written as
\begin{align}
\frac{2 (c_{Hq} - c'_{Hq})}{\Lambda^2} = \left. \frac{d \tilde{\A}_{u_L \, \phi^{0}}}{d s} \right|_{s=0}    =&~ \int^\infty_{0}  \frac{ds}{\pi s}  \left( \sigma_{\rm tot}^{u_L \, \phi^{0} } - \sigma_{\rm tot}^{u_L \, \phi^{0*} }  \right) + c_\infty \,,  \nonumber\\ 
 \frac{2 c_{Hu}}{\Lambda^2} = \left. \frac{d \tilde{\A}_{u_R \, \phi^0}}{d s} \right|_{s=0}    =&~ \int^\infty_{0}  \frac{ds}{\pi s}  \left( \sigma_{\rm tot}^{u_R \, \phi^0 } - \sigma_{\rm tot}^{u_R \, \phi^{0*} }  \right) + c_\infty \,, \nonumber\\
 \frac{2 (c_{Hq} + c'_{Hq})}{\Lambda^2} = \left. \frac{d \tilde{\A}_{d_L \, \phi^0}}{d s} \right|_{s=0}    =&~ \int^\infty_{0}  \frac{ds}{\pi s}  \left( \sigma_{\rm tot}^{d_L \, \phi^0 } - \sigma_{\rm tot}^{d_L \, \phi^{0*} }  \right) + c_\infty \,,  \nonumber\\
 \frac{2 c_{Hd}}{\Lambda^2} = \left. \frac{d \tilde{\A}_{d_R \, \phi^0}}{d s} \right|_{s=0}    =&~ \int^\infty_{0}  \frac{ds}{\pi s}  \left( \sigma_{\rm tot}^{d_R \, \phi^0  } - \sigma_{\rm tot}^{d_R \, \phi^{0*}  }  \right) + c_\infty \,,   \label{eq:sumqh1} 
\end{align}
where we have picked up the neutral component of Higgs doublet. Equivalently, one could written down 4 equations with the charged component of the Higgs, related to \autoref{eq:sumqh1} by an $SU(2)_L$ rotation ($u_L \leftrightarrow d_L$, $\phi^+ \leftrightarrow \phi^0$, $\phi^- \leftrightarrow \phi^{0*}$).  For the leptons, the 3 sum rules can be written as
\begin{align}
\frac{2 (c_{Hl} - c'_{Hl})}{\Lambda^2} = \left. \frac{d \tilde{\A}_{\nu_L \, \phi^{0}}}{d s} \right|_{s=0}    =&~ \int^\infty_{0}  \frac{ds}{\pi s}  \left( \sigma_{\rm tot}^{\nu_L \, \phi^{0} } - \sigma_{\rm tot}^{\nu_L \, \phi^{0*} }  \right) + c_\infty \,,  \nonumber\\ 
 \frac{2 (c_{Hl} + c'_{Hl})}{\Lambda^2} = \left. \frac{d \tilde{\A}_{e_L \, \phi^0}}{d s} \right|_{s=0}    =&~ \int^\infty_{0}  \frac{ds}{\pi s}  \left( \sigma_{\rm tot}^{e_L \, \phi^0 } - \sigma_{\rm tot}^{e_L \, \phi^{0*} }  \right) + c_\infty \,,  \nonumber\\
 \frac{2 c_{He}}{\Lambda^2} = \left. \frac{d \tilde{\A}_{e_R \, \phi^0}}{d s} \right|_{s=0}    =&~ \int^\infty_{0}  \frac{ds}{\pi s}  \left( \sigma_{\rm tot}^{e_R \, \phi^0  } - \sigma_{\rm tot}^{e_R \, \phi^{0*}  }  \right) + c_\infty \,,   \label{eq:sumlh1} 
\end{align}
with the absent one corresponding to the lack of $\nu_R$ in the SM.
We note again that the operators in \autoref{eq:ohw} also contribute to the Higgs-fermion amplitudes but can be exchanged to the $\mathcal{O}_{Hf}$ type operators and do not have additional independent contributions.  

%%%%%%%%%%%%%%%%%%%%%%%%%%%%%%%
\subsection{Fermion-fermion amplitudes}

The elastic 4-fermion amplitudes in the SM can be obtained by scattering any two of the five fermion fields $f=\, q_L, \, l_L, \,  u_R, \, d_R, \,e_R$.  There are 15 combinations in total, 5 from scattering two identical fermions and 10 from two different fermions.  Among them, we find that the following five combinations each contains two independent ways of contracting group indices (with $i,j$ denoting $SU(2)$ indices and $a,b$ denoting $SU(3)$ indices):

\begin{itemize}

  \item  $\A(qq\bar{q}\bar{q})$:  two independent amplitudes can be obtained from two ways of contracting $SU(2)$ and $SU(3)$ indices, which are $\A(q^a_i q^b_j \bar{q}^a_i \bar{q}^b_j)$ and  $\A(q^a_i q^b_j \bar{q}^b_i \bar{q}^a_j)$;
  
  \item  $\A(lq\bar{l}\bar{q})$: one could contract the SU(2) indices between the two leptons (quarks) or between one lepton and one quark, giving $\A(l_i q^a_j \bar{l}_i \bar{q}^a_j)$ and $\A(l_i q^a_i \bar{l}_j \bar{q}^a_j)$;
  
  \item  $\A(qu\bar{q}\bar{u})$, $\A(qd\bar{q}\bar{d})$ and $\A(ud\bar{u}\bar{d})$:  in each case one could contract the $SU(3)$ indices between the same fermion or between different ones;

\end{itemize}
while the other combinations each contains one independent amplitude.  As such, a total number of 20 independent amplitudes can be written down.  Each amplitude contain one parameter, as the kinematic structure is fixed as shown in \autoref{tab:a4elastic}.  By considering the scattering of different states ({\it e.g.} setting $i=j$ and/or $a=b$), a total number of 20 sum rules can be written down for the 20 parameters.

Our counting matches the number of 4-fermion operators in the SMEFT, excluding those composed of 4 different fermions (which could not contribute to elastic amplitudes).  These 20 operators can be found in {\it e.g.} Ref.~\cite{Grzadkowski:2010es} under the $(\bar{L}L)(\bar{L}L)$, $(\bar{R}R)(\bar{R}R)$ and $(\bar{L}L)(\bar{R}R)$ categories.  Not surprisingly, all of these 20 operators have only the $\psi^+ \psi^+ \psi^- \psi^-$ helicity configuration, and contribute to the elastic amplitudes in the case of one fermion generation.

Due to the large number of sum rules,  we will only show one example, $e_R \, \overline{e_R} \to e_R \, \overline{e_R}$, generated by the 4-fermion interaction term $ \frac{c_{ee}}{\Lambda^2} (\overline{e_R} \gamma_\mu e_R)(\overline{e_R} \gamma^\mu e_R)$.  Its sum rule can be written as
\begin{equation}
-\frac{2 c_{ee}}{\Lambda^2} =  \left. \frac{d \tilde{\A}_{e_R \, \overline{e_R} }}{d s} \right|_{s=0}    = \int^\infty_{0} \frac{ds}{\pi s} \left( \sigma^{e_R \, \overline{e_R} }_{\rm tot} - \sigma^{e_R\, e_R }_{\rm tot}  \right) + c_\infty \,.  \label{eq:sumee1} 
\end{equation}
%

%%%%%%%%%%%%%%%%%%%%%%%%%%%%%

\subsection{Comparison with previous results}

Sum rules of dimension-6 operators have also been examined in previous literatures~\cite{Low:2009di, Falkowski:2012vh, Bellazzini:2014waa}.    
Our study is distinguished from them by focusing on the amplitudes in the unbroken electroweak phase, which makes the connection between the amplitudes and the dimension-6 operators more transparent.   Some of our results are nevertheless closely connected to the previous ones.  In particular, \autoref{eq:a4hO1} was already pointed out in Refs.~\cite{Low:2009di, Falkowski:2012vh} in the context of the Goldstone fields, which has the same implications on the operator coefficient $c_H$ as we discuss later in \autoref{sec:bm}.  Sum rules for new physics with additional global symmetries (such as composite Higgs models) has been throughly discussed in Ref.~\cite{Bellazzini:2014waa}.  Many previous studies also focuses on the applications in chiral perturbation theory and nuclear theory, including pion-nucleon scatterings \cite{Goldberger:1955zza, Hamilton:1963zz, Luo:2006yg, Sanz-Cillero:2013ipa} and fermion scatterings \cite{Adams:2008hp}.  We focus on the SMEFT and do not consider those cases.

%%%%%%%%%%%%%%%%%%%%%%%%%%%%%%%%%%%%%%%%%%%%%%%%%%
\section{Implications of the sum rules}
\label{sec:imp}
%%%%%%%%%%%%%%%%%%%%%%%%%%%%%%%%%%%%%%%%%%%%%%%%%%

\subsection{Robustness of the sum rules}
\label{sec:valid}

Each of the sum rules in Eqs.\,(\ref{eq:sumhh1}--\ref{eq:sumee1}) can be interpreted as a relation between the EFT on the lefthand side and the quantities in the full theory on the righthand side.  The SM contributions are assumed to be absent in these sum rules, as they only contribute to the $\A^{[0]}_4$ term in \autoref{eq:ampexp} in the massless limit.  Here we discuss the possible caveats of this assumption and show that even in these cases, the presence of SM contributions does not obscure the interpretation of the sum rules.  
As mentioned earlier, considering an energy scale sufficiently higher than the electroweak scale, $\mu^2\gg m^2$, it is reasonable to treat the SM particles as being approximately massless. In this limit, 
the SM could not generate poles in the forward amplitudes.  A divergent forward amplitude can be generated by $t$-channel diagrams ({\it e.g.} of a photon) which contribute to the boundary term.  This contribution can thus be subtracted from both side of the sum rule without any impact on the physics implication.  If the SM particle masses are restored, an $s$ or $u$-channel exchange of a SM particle would then have corresponding poles in the $s$-plane, thereby giving a contribution to the righthand side of the sum rules, either to the cross section terms in \autoref{eq:A03}, or as additional IR poles if the mass is smaller than $2m$.    
They also modify the $c_n$ on the lefthand side.  These contributions have to match, and can be computed on both side and subtracted from the sum rule.  Similarly, the SM loop contribution to the forward amplitudes matches the $2\to 2$ SM cross sections.  Different from tree level contributions which are low energy poles, the loop contributions are branch cuts in the $s$-plane and extend to high energies.  Nevertheless, their contribution  usually dominates at low energies, for which the SMEFT is valid and the SM contributions are calculable and can be subtracted~\cite{Bi:2019phv}.  
Loop corrections from massless particles may also contain IR divergence, in which case an IR cutoff in the dispersion relation ({\it i.e.} restoring the mass $m$ in \autoref{eq:A03}) is required to regulate this contribution.  We assume this is always done before subtracting the loop contributions of massless SM particles.

Next, we comment on the robustness of the sum rules under one-loop effects of new physics.  
From a pure EFT point of view, to include such contributions in the low energy measurement, we should first perform a one-loop matching. The resulting operators will then be evolved to the scale appropriate for the measurement using renormalization group equations (RGE)~\cite{Grojean:2013kd, Jenkins:2013zja,Jenkins:2013wua,Alonso:2013hga, Elias-Miro:2013gya, Elias-Miro:2013mua}. 
Recently, a deeper connection has been established between the helicity amplitude structures of SMEFT and the RGEs of the operator coefficients~\cite{Craig:2019wmo, Cheung:2015aba, Azatov:2016sqh, Bern:2019wie, Jiang:2020sdh, Bern:2020ikv, Baratella:2020lzz,  EliasMiro:2020tdv, Jiang:2020mhe}.  In short, an operator is only renormalized by another if the latter contributes at one loop a divergent contribution to the helicity amplitude which corresponds to a contact interaction of the former.
From this point of view, the sum rules of the helicity amplitudes should be able to capture the operator mixing effects, if the one-loop contributions of the new physics model are included and if the corresponding operators are generated by the model.  As such, we expect the sum rules to reproduce the results of one-loop matching and RG running of the SMEFT.  This is verified specifically for an example model later in \autoref{sec:bemi}.

%%%%%%%%%%%%%%%%%%%%%%%%%%%%%%%%%%%%%%
\subsection{Custodial symmetries}
\label{sec:sym}
%%%%%%%%%%%%%%%%%%%%%%%%%%%%%%%%%%%%%%

The sum rule in \autoref{eq:sum1} suggests that the amplitude  $\tilde{\A}^{[2]}_{ab}(s) = \left. \frac{d \tilde{\A}_{ab}(s)}{d s}  \right|_{s=0} $ could be highly suppressed if $\sigma^{ab}_{\rm tot} \approx \sigma^{a\bar{b}}_{\rm tot}$.  It is possible that the full theory exhibits certain (at least approximate) symmetries which fulfill this condition without fine tuning.  We find that such symmetries can indeed be imposed in a general sense regardless of the boundary term, and leads to the familiar custodial symmetries for the Higgs-Higgs and Higgs-fermion amplitudes.  
To start, we recall that in the massless limit, $\tilde{\A}^{[2]}_{ab}(s) \propto s$ and is an odd function of it.  Under the $s \leftrightarrow u$ crossing one has
\begin{equation}
\tilde{\A}^{[2]}_{ab}(s) \underset{s \leftrightarrow u}= \tilde{\A}^{[2]}_{a\bar{b}} (u) = \tilde{\A}^{[2]}_{a\bar{b}}(-s)= - \tilde{\A}^{[2]}_{a\bar{b}}(s)  \,.
\end{equation}
To make $\tilde{\A}^{[2]}_{ab}(s)$ vanish, one simply needs to make it also an even function of $s$.  This can be done by imposing a symmetry (denoted as $\sym$) on the theory so that it is invariant under the mapping:
\begin{equation}
\sym: \hspace{0.5cm} a \to  a  \,, \hspace{0.5cm}   b \to \bar{b}\,.  \label{eq:symab}
\end{equation}
To summarize, we have
\begin{align}
\tilde{\A}^{[2]}_{ab} (s)=&~ - \tilde{\A}^{[2]}_{a\bar{b}} (s)   \hspace{1cm}  \mbox{under crossing:  } s\leftrightarrow u \,, \label{eq:sym1} \\
\tilde{\A}^{[2]}_{ab}(s) =&~ \tilde{\A}^{[2]}_{a\bar{b}}(s)    \hspace{1.4cm}  \mbox{~under } \sym:  a\leftrightarrow a\,,~~b\leftrightarrow \bar{b} \,, \label{eq:sym2}
\end{align}
and combining the two equations we arrive at the result that $\tilde{\A}^{[2]}_{ab} = 0$.  
Denoting 
some common quantum number (with label $i$) of particles $a$ and $b$ as $\sigma^i_a$ and $\sigma^i_b$, we require that the theory must be invariant under
\begin{equation}
\sigma^i_a \to  \sigma^i_a  \,, \hspace{1cm}   \sigma^i_b \to -\sigma^i_b\,,  \label{eq:ab}
\end{equation}
for all $i$s. The condition in \autoref{eq:ab} is generally nontrivial and can be satisfied in two ways:  
\begin{enumerate}

\item  For each nonzero $\sigma^i_a$, one has $\sigma^i_b = 0$, and vice versa.  In this case $\sym$ is either trivial or an overall transformation that makes $\sigma^i \to - \sigma^i$. 

\item  Under certain setups, $\sym$ can be a parity that exchanges some of the quantum numbers.  For example, suppose $i=1,2$, $\sigma ^1_a = \sigma^2_a$ and $\sigma^1_b = -\sigma^2_b$, then a parity that exchanges the two quantum numbers $P(1\leftrightarrow 2)$ satisfies \autoref{eq:ab}.

\end{enumerate}
Note that the two conditions above are also symmetric under $a$ and $b$.  One could equivalently consider the crossing of particle $a$ and require the symmetry $\sym'$ that
\begin{equation}
\sym': \hspace{0.5cm} a \to  \bar{a}  \,, \hspace{0.5cm}   b \to b \,.
\end{equation}
$\sym$ and $\sym'$ can be exchanged by a $CP$ transformation along with a spatial rotation 
to compensate $\vec{p} \to - \vec{p}$ from parity. 
The massless forward elastic amplitudes are indeed invariant under $CP$, since one gets the original amplitude by crossing it twice ($\tilde{\A}_{ab} \to \tilde{\A}_{a\bar{b}} \to \tilde{\A}_{\bar{a}\bar{b}} $).\footnote{Note that this is only true in the helicity basis in which the particle polarizations are invariant under spatial rotations.  In the linear basis, for instance, the 4-vector elastic amplitude receives contribution from the CP-odd dimension-8 operators~\cite{Remmen:2019cyz}.}
It is also clear that $\sym$ or $\sym'$ could not eliminate the $\tilde{\A}^{[4]}_{ab}$ amplitudes (generated by dimension-8 operators), which instead exhibit positivity relations.
The SM Higgs sector has a $SU(2)_L \times SU(2)_R$ global symmetry.  Naturally, a parity that relates the left-handed and right-handed symmetries, $P_{LR}$, could fulfill the requirement of $\sym$.  Indeed, the embedding of the Higgs doublet under $SU(2)_L \times SU(2)_R$ gives the following quantum numbers
\begin{equation}
\bordermatrix{ 
~~_{t_{3L}} \!\! \backslash \!\!\! ~^{t_{3R}} & ~^{1/2}   &  ~^{-1/2}  \cr 
~~_{1/2} & \phi^+ &  \phi^{0*} \cr
~_{-1/2} & \phi^0 &  -\phi^-
}\,, %_0 \,, 
 \hspace{1cm} \mbox{where}~
H= \bpm \phi^+ \\ \phi^0 \epm \,.
\end{equation}
Under a $P_{LR}$ symmetry that exchanges $SU(2)_L$ and $SU(2)_R$, we have
\begin{equation}
P_{LR}: \hspace{0.5cm} \phi^+ \to \phi^+, \hspace{0.5cm} \phi^0 \to \phi^{0*}\,,
\end{equation}
which makes the following amplitude vanish,
\begin{equation}
\tilde{\A}^{[2]}_{\phi^+ \phi^0}  \equiv \A^{[2]} (\phi^+ \phi^0 \to  \phi^+ \phi^0)|_{t=0}  = -\frac{2c_T}{\Lambda^2}s  = 0 \,.  \label{eq:symcus} 
\end{equation}
Indeed, this is nothing but the consequence of the custodial symmetry.  The operator $O_T$ breaks the custodial symmetry together with $P_{LR}$, so $c_T$ must vanish if the symmetry is preserved.  A more intuitive understanding can be obtained by using $SO(4) \sim SU(2)_L \times SU(2)_R$ and writing the Higgs doublet in terms of the field $\Phi$ with four real components, 
\begin{equation}
H=\frac{1}{\sqrt{2}} \bpm \phi_2 + i \phi_1 \\ \phi_4 + i\phi_3 \epm \,,  \hspace{1cm} \Phi = \bpm \phi_1 \\  \phi_2 \\  \phi_3 \\  \phi_4 \epm \,. 
\end{equation}
We could then enlarge the symmetry group to $O(4)$ by imposing a parity that flips the sign of any of the $\phi_i$s.  In fact, with only the SM Higgs field, it is not possible to write down a term that preserves $SO(4)$ while breaking the parity.  This parity exchanges either $\phi^0 \leftrightarrow \phi^{0*}$ or $\phi^+ \leftrightarrow \phi^-$, and is exactly the symmetry needed for \autoref{eq:symcus} to hold.  On the other hand, with $P_{LR}$ it is not possible to explicitly break $SU(2)_R$ without also breaking $SU(2)_L$ and violate gauge invariance.  We thus conclude that \autoref{eq:symcus} holds if and only if the $SU(2)_L \times SU(2)_R$ symmetry of the SM Higgs sector is preserved.

For the Higgs-fermions amplitudes, the same symmetry can be imposed with certain embedding of the fermion $f$ under $SU(2)_L \times SU(2)_R$.  The symmetry
\begin{equation}
P_{LR}: \hspace{0.5cm} f \to f, \hspace{0.5cm} \phi^0 \to \phi^{0*}\,,
\end{equation}
can be imposed by requiring the isospins of $f$ to satisfy either  
\begin{equation}
T^3_L = T^3_R = 0 \,, \hspace{1cm}{\mbox or}\hspace{1cm}  T_L=T_R \,, ~~ T^3_L = T^3_R\,,  \label{eq:conzbb}
\end{equation}
which are exactly the same conditions in Ref.~\cite{Agashe:2006at} for protecting the SM $Zf\bar{f}$ coupling.  In particular, the second condition in \autoref{eq:conzbb} is very common in the construction in composite Higgs models for protecting the $Zb_L\bar{b}_L$ coupling, which we discuss further in \autoref{sec:zbb}.

In principle one could also apply the symmetry to the fermion-fermion amplitudes.  However, note the symmetry $\sym$ in \autoref{eq:symab} also flips the helicity of particle $b$.  For the the fermion-fermion amplitude, it can be shown that the two amplitudes $\A^{[2]}_{ab}$ and $\A^{[2]}_{a\bar{b}}$ have different total angular momenta~\cite{Jiang:2020sdh}.  At tree level, this means that the symmetry necessarily relates heavy particles with different spins.
More specifically, consider the $2\to 1$ cross section of two fermions to a heavy scalar,
and the two fermions must have the same helicity.
By changing one of the fermion to its antiparticle, they will have opposite helicity and the final state must be a vector.  The symmetry thus connects a scalar with a vector.  We do not consider such possibilities in this paper.

\subsection{Boundary term}
The boundary term, $c_\infty =  \frac{1}{2\pi i} \oint\limits_{s\to \infty} ds \frac{\tilde{\A}(s)}{s^2}$, is generally nonzero and needs to be included in the sum rule.  
The typical contribution from a weakly coupled UV theory at tree level is from the $t$-channel exchange of a heavy vector.  Assuming the low energy forward amplitude $\tilde{\A}(s) = \frac{g}{M^2} s + ...$ is generated by such a $t$-channel diagram, one must have in the full theory  
\begin{equation}
\tilde{\A}(s) \rightarrow \left. \frac{-g\, s}{t-M^2} \right|_{t=0} = \frac{g\,s}{M^2} 
  \hspace{0.5cm} \Rightarrow   \hspace{0.5cm}
c_\infty = \frac{g}{M^2}   \,,  \label{eq:auvinf}
\end{equation}
such that $\left. \frac{d \tilde{\A}}{d s} \right|_{s=0} = c_\infty$.  The presence of the boundary term could obscure the connection between the EFT parameters and the cross section terms.  In many scenarios of interest, this $t$-channel contribution is absent and $c_\infty$ can be set to zero.  In strongly coupled UV scenarios, the boundary terms have also been shown to vanish under generic conditions~\cite{Falkowski:2012vh, Bellazzini:2014waa}. 
However, it should be emphasized that the properties of $c_\infty$ are model dependent and needs to be treated with caution.  

We note in \autoref{eq:auvinf} that the boundary term has the same $s\leftrightarrow u$ cross relation as $\tilde{\A}^{[2]}_4$, so the symmetry in \autoref{eq:sym2} also makes the boundary term vanish.   This has to be the case since the symmetry would make everything else in the sum rule vanish, and the boundary term must therefore vanish as well.  
As an example, we consider a single heavy vector $Z'$ that couples to the SM Higgs and ignore the SM gauge bosons for simplicity, %
\begin{align}
\mathcal{L} =&~ -\frac{1}{4} Z'_{\mu\nu}  Z'^{\mu\nu} - |D_\mu H|^2 + ...  \nonumber \\
=&~ ... + i g_{Z'} Z'^\mu [ (\partial_\mu H^\dagger )H -  H^\dagger \partial_\mu H ] +...   \label{eq:Lsqed}
\end{align}
where the vector-scalar-scalar vertex has a coupling in the form $i g_{Z'} (p_2-p_1)^\mu$.   
Looking at its contribution to $\A^{[2]} (\phi^+ \phi^0 \to  \phi^+ \phi^0)$ in \autoref{eq:a4hO2}, 
it will generate a term in the form $\frac{s-u}{t-M^2}$, and is indeed odd under the $s \leftrightarrow u$ crossing.  More over, upon integrating out the $Z'$, we will generate an operator in the form $(H^\dagger \overleftrightarrow{\partial_\mu} H)^2$, which after adding the SM gauge bosons is just $\mathcal{O}_T$ which breaks the custodial symmetry, in agreement with \autoref{eq:sumhh2}.

%%%%%%%%%%%%%%%%%%%%%%%%%%%%%%%%%%%%%%%%%%%%%%%%%%%%%%%%%%%%%%%
\subsection{Precision measurements and direct searches}    
The sum rules in Eqs.\,(\ref{eq:sumhh1}--\ref{eq:sumee1}) establish the general connections between the SMEFT and properties of the full model.  The SMEFT can be probed by precision measurements at low energy where the leading order contributions from new physics are parameterized by the dimension-6 Wilson coefficients. 
While the sum rules are obtained in the massless SMEFT without the Higgs vev, our results on the Wilson coefficients can nevertheless be connected to observables around the electroweak scale, to which the contributions from dimension-6 operators are known.  In particular, $c_T$ is related to the $T$-parameter~\cite{Peskin:1991sw} that can be constrained by the $Z$-pole and $W$ mass measurements, while $c_H$ can be probed by the measurements of Higgs couplings. 
The $c_{Hf}$ parameters modify the couplings of the fermions to the weak gauge bosons.    
Each sum rule in \autoref{eq:sumqh1} and \autoref{eq:sumlh1} can be connected to the modification of the couplings of fermions to the $Z$ boson after the electroweak symmetry breaking.  More specifically, with the parameterization
\begin{equation}
\mathcal{L} = \frac{g}{c_W} Z_\mu \left( \underset{f=u,d,\nu,e}{\sum} \bar{f}_L \gamma^\mu (T_3 -s^2_W Q + \delta g_{Lf}) f_L + \underset{f=u,d,e}{\sum} \bar{f}_R \gamma^\mu ( -s^2_W Q + \delta g_{Rf}) f_R  \right) + ... \,,  \label{eq:Lzff}
\end{equation}
where $s^2_W$ and $c^2_W$ are shorthands for $\sin^2 \theta_W$ and  $\cos^2 \theta_W$ and $\theta_W$ is the weak mixing angle, the 7 equations in \autoref{eq:sumqh1} and \autoref{eq:sumlh1} equal to %
$\frac{4}{v^2} \times \{g_{Lu},\,g_{Ru},\,g_{Ld},\,g_{Rd},\,g_{L\nu},\,g_{Le},\,g_{Re} \}$,
respectively.  The scattering processes at high energies (with $v \ll E \ll \Lambda$), such as the Higgsstrahlung process at hadron or lepton colliders ($pp \to Vh$ and $e^+ e^- \to Zh$), offer a more direct probe of the corresponding amplitudes.  
The properties of the full model, on the other hand, can be probed by direct searches at high energies.  Different from the EFT parameters that are subject to the sum rules, the direct search bound can be applied to individual particles, and are thus complementary to the bounds from precision measurements.    
\begin{figure}[t]
\centering
\includegraphics[width=0.5\textwidth]{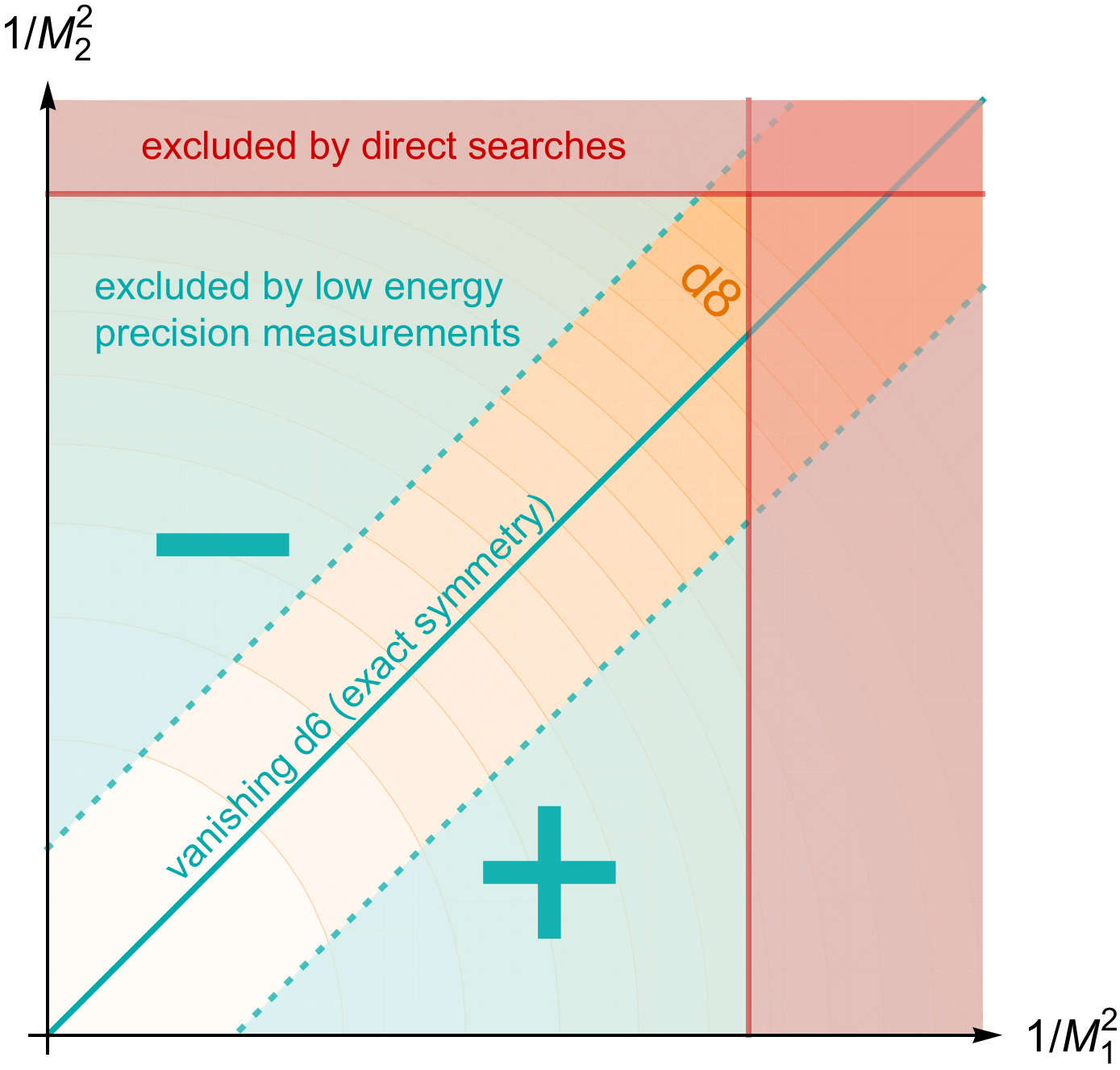} 
\caption{A schematic plot on the interplay between precision measurements and direct searches.  For simplicity, we assume only two new particles $X_1$ and $X_2$ with masses $M_1$ and $M_2$ and some universal couplings to SM.  They each contribute to one of the cross sections in the sum rule, with $\sigma(ab\to X_1)$ and $\sigma(a\bar{b} \to X_2)$.   The symmetry in \autoref{eq:symab} corresponds to the diagonal line, where the contribution to $\A(ab\to ab)$ from dimension-6 operators vanishes, while the plus (minus) sign denotes the region in which this contribution is positive (negative).  Contributions to $\A(ab\to ab)$ from dimension-8 operators are proportional to $\frac{1}{M^4_1} + \frac{1}{M^4_2}$, as illustrated by the orange contours.  
}
\label{fig:schem}
\end{figure}
This complementarity is illustrated schematically in \autoref{fig:schem}.  
For simplicity, we assume in \autoref{fig:schem} that the full theory contains only two heavy particles, $X_1$ and $X_2$, with masses $M_1$ and $M_2$ and some universal couplings to SM.  They contribute to the cross sections $\sigma(ab\to X_1)$ and $\sigma(a\bar{b} \to X_2)$ in the sum rule for the forward amplitude $\tilde{\A}(ab\to ab)$.  As such, the contributions to $\tilde{\A}(ab\to ab)$ from dimension-6 operators are proportional to $\frac{1}{M^2_1} - \frac{1}{M^2_2}$, corresponding to diagonal lines in the $(\frac{1}{M^2_1}, \frac{1}{M^2_2})$ plane.  They can be constrained by low energy precision measurements.  The direct searches for $X_1$ and $X_2$ are mostly independent of each other, resulting in a rectangular allowed region.

It is interesting to note that the symmetry in \autoref{eq:symab} also provides a plausible scenario where dimension-8 operators could dominate certain scattering processes.   The contributions to $\tilde{\A}(ab\to ab)$ from dimension-8 operators are proportional to $\frac{1}{M^4_1} + \frac{1}{M^4_2}$, represented by the orange circular contours in \autoref{fig:schem}.  Near the region $M_1 \approx M_2$, the dimension-8 operators could give the dominant contribution to $\tilde{\A}(ab\to ab)$.  
An approximate symmetry that suppresses $\frac{1}{M^2_1} - \frac{1}{M^2_2}$ naturally constrains the parameter space to be in this region.  The symmetry can also be explicitly tested by precision measurements at low energy ({\it e.g.} at $Z$-pole) for which the contributions from dimension-8 operators are highly suppressed.  On the other hand, the measurements of scattering process at high energy (such as $pp\to Vh$ or $e^+e^- \to Zh$) could be used to probe the dimension-8 operators, and test the positivity bound associated with them.
We also note that the same symmetry also suppresses the contribution from dimension-10 operators, making the validity of the EFT expansion more robust for those high energy measurements.  
A more detailed analysis of this interesting scenario is left for future studies.

%%%%%%%%%%%%%%%%%%%%%%%%%%%%%%%%%%%%%%
\section{Benchmark models}
\label{sec:bm}
%%%%%%%%%%%%%%%%%%%%%%%%%

In this section, we demonstrate the applications of the sum rules, using several new physics models as examples.

%%%%%%%%%%%%%%%%%%%%%%%%%
\subsection{Doubly charged scalars}
%%%%%%%%%%%%%%%%%%%%%%%%%

Since the two cross section terms in the sum rules have opposite signs, they contain useful information on the possible signs of the Wilson coefficients.  As a first example, let us take the Higgs sum rule in \autoref{eq:sumhh1}, and assume that $c_T=0$ due to the strong constraints from electroweak measurements.  We could then write
\begin{equation}
\frac{c_H }{\Lambda^2}  = \int^\infty_{0} \frac{ds}{\pi  s} \left( \sigma^{\phi^+ \phi^-}_{\rm tot} - \sigma^{\phi^+ \phi^+}_{\rm tot}   \right) + c_\infty \,. \label{eq:sumhh1x}
\end{equation}
The boundary term $c_\infty$ is generated only by $t$-channel vector boson exchanges at tree level.  One could show that $c_\infty>0$ since it is proportional to the square of the Higgs-Higgs-vector coupling.
A negative $c_H$ can thus only be generated by a charge-$2$ scalar (which contributes to $ \sigma^{\phi^+ \phi^+}_{\rm tot}$) at tree level~\cite{Low:2009di}.  A similar argument can also be applied to \autoref{eq:sumee1}, which implies that a charge-$2$ scalar is also needed to generate a positive $c_{ee}$ at tree level.

%%%%%%%%%%%%%%%%%%%%%
\subsection{Triplet scalars}
\label{sec:tris}
%%%%%%%%%%%%%%%%%%%%%

Models with heavy triplet scalars provide an interesting case for the Higgs sum rules in \autoref{eq:sumhh1} and \autoref{eq:sumhh2} since they can contribute to both $c_H$ and $c_T$.   
At tree level, the contributions to the Higgs 4-point amplitudes must come from an intermediate heavy scalar.  For the sum rules it is thus sufficient to consider only the 3-scalar interactions of two Higgs and one heavy scalar.  In this case, the hypercharge of the triplet scalar needs to be either $0$ or $\pm 1$.  The relevant interaction terms in the Lagrangian can be written as
\begin{equation}
\mathcal{L}_{\rm int} = \kappa_\xi H^\dagger \sigma^a H \xi_a + \frac{\kappa_\chi}{\sqrt{2}} ( \widetilde{H}^\dagger \sigma^a H \chi_a + {\rm h.c.}) \,,  \label{eq:ts1}
\end{equation}
where $\xi$ and $\chi$ are triplet scalars with hypercharge $0$ and $-1$, respectively.  The couplings $\kappa_\xi$ and $\kappa_\chi$ have mass dimension one.  Writing down the components explicitly, one has
\begin{align}
\mathcal{L}_{\rm int} =&~~~ \kappa_\xi \left[ \sqrt{2}\, \phi^0 \phi^- \xi^+ +  \sqrt{2} \, \phi^{0*} \phi^+ \xi^- +  (\phi^0 \phi^{0*} - \phi^- \phi^+ ) \xi^0  \right] \nonumber \\
&+ \kappa_\chi \left[ \phi^0 \phi^0 \chi^{0*} + \phi^+ \phi^+ \chi^{--} + \sqrt{2}\, \phi^0 \phi^+ \chi^-  + {\rm h.c.}   \right] \,,
\end{align}
where
\begin{align}
\xi^+ =&~ \frac{\xi_1-i\xi_2}{\sqrt{2}} \,,  \hspace{1.3cm}   \xi^0 = - \xi_3   \,,   \hspace{1cm}  \xi^- = \frac{\xi_1+ i\xi_2}{\sqrt{2}} \,,   \nonumber \\
\chi^{0*} =&~ \frac{\chi_1-i\chi_2}{\sqrt{2}} \,,  \hspace{1cm}  \chi^- = \chi_3\,,  \hspace{1cm}   \chi^{--} = -\frac{\chi_1+ i\chi_2}{\sqrt{2}} \,,
\end{align}
with all fields canonically normalized.
The triplet interactions in \autoref{eq:ts1} do not contribute to the boundary term $c_\infty$ in the sum rules. This is because the $t$-channel amplitudes have the form $\sim \frac{\kappa^2}{t-m^2}$ and have no $s$ dependence.  Therefore, we could set $c_\infty$ to zero and write \autoref{eq:sumhh1} and \autoref{eq:sumhh2} as\footnote{Note that the $\phi^+ \phi^+ \chi^{--}$ vertex has a symmetry factor of 2, which gives the extra factor of 4 in $\sigma(\phi^+ \phi^+ \to \chi^{++})$. }
\begin{align}
\frac{c_H + 3c_T}{\Lambda^2} = &~ \int^\infty_{0} \frac{ds}{\pi  s} \left( \sigma^{\phi^+ \phi^- \to \xi^0 } - \sigma^{\phi^+ \phi^+ \to \chi^{++}}   \right) =  \frac{\kappa^2_\xi}{m^4_\xi} - \frac{ 4 \kappa^2_\chi}{m^4_\chi}   \,, \label{eq:sumts1} \\
- \frac{2 c_T}{\Lambda^2} = &~   \int^\infty_{0}  \frac{ds}{\pi s} \left( \sigma^{\phi^+ \phi^0 \to \chi^+} - \sigma^{\phi^+ \phi^{0*} \to \xi^+}  \right)  = \frac{2\kappa^2_\chi}{m^4_\chi} - \frac{2\kappa^2_\xi}{m^4_\xi} \,, \label{eq:sumts2}
\end{align}
where $m_\xi$ ($m_\chi$) is the mass of $\xi$ ($\chi$).  Note that $\kappa_\xi$ and $\kappa_\chi$ both have mass dimension one, and $\kappa^2/m^4 \sim 1/\Lambda^2$ as expected. 

As shown in \autoref{eq:sumts2}, the two triplet scalars in \autoref{eq:ts1} both contribute to $c_T$, but with opposite signs.  It is thus possible to arrange cancellations of the two terms by imposing the custodial symmetry, as in the Georgi-Machacek Model~\cite{Georgi:1985nv}.  Indeed, by setting $\kappa=\kappa_\xi=\kappa_\chi$ and $m = m_\xi = m_\chi $ we reproduce the trilinear interactions in the Georgi-Machacek Model in Refs.~\cite{Aoki:2007ah, Chiang:2012cn, Hartling:2014zca}.\footnote{Our conventions for the fields are also chosen to match the ones in these references.  In particular, by setting $\kappa_\xi = \kappa_\chi \to -M_1/2$ and $m^2_\xi = m^2_\chi \to \mu^2_3$ we reproduce the trilinear term $-M_1 {\rm Tr}(\Phi^\dagger \tau^a \Phi \tau^b) (UXU^\dagger)_{ab}$ in Eq.~(5) of Ref.~\cite{Hartling:2014zca}. }
On the other hand, we see from \autoref{eq:sumts1} that $c_H$ receives a negative overall contribution in this limit, $\frac{c_H}{\Lambda^2} = -3 \frac{\kappa^2}{m^4}$.  
Our result for $c_H$ agrees with the one in  Ref.~\cite{Hartling:2014zca} (expressed in terms of $\kappa_V = 1 - \frac{c_H}{2}\frac{v^2}{\Lambda^2}$), but is obtained with much less effort with the help of the sum rules.\footnote{To be precise, both $\mathcal{O}_H$ and $\mathcal{O}_r = |H|^2|D_\mu H|^2$ are generated in the Georgi-Machacek model.  Their contributions to the 4-scalar amplitude could not be distinguished, but $\mathcal{O}_r$ also contributes to the $hhhhVV$ contact interaction.  $\mathcal{O}_H$ modifies the Higgs couplings universally, while $\mathcal{O}_r$ only modifies the couplings to gauge bosons.  $\mathcal{O}_r $ is usually eliminated via field redefinition, and can be replaced by a combination of $\mathcal{O}_H$, $\mathcal{O}_6$, and $\mathcal{O}_y$ operators which directly modify the Yukawa couplings~\cite{Elias-Miro:2013mua}.  This explains why $\kappa_V$ and $\kappa_f$ are different in Ref.~\cite{Hartling:2014zca}.  }

It is also interesting to note that by combining \autoref{eq:sumts1} and \autoref{eq:sumts2}, we obtain
\begin{equation}
\frac{c_H}{\Lambda^2} =  - \frac{2 \kappa^2_\xi}{m^4_\xi} - \frac{  \kappa^2_\chi}{m^4_\chi}   \,,
\end{equation}
suggesting that $c_H<0$ for any triplet scalar extension, even for the $\xi$ triplet scalar alone which does not contain a doubly charged scalar.  Needless to say, $\xi$ alone gives a nonzero $c_T$ and is strongly disfavored by electroweak measurements.

%%%%%%%%%%%%%%%%%%%%%%%%%%%%%%%%%%%%%%
\subsection{The Beautiful Mirror model}
\label{sec:bemi}
%%%%%%%%%%%%%%%%%%%%%%
The Beautiful Mirror (BM) model, proposed in Ref~\cite{Choudhury:2001hs}, provides an interesting benchmark for both the Higgs-fermion and the Higgs-Higgs sum rules.  The BM model introduces exotic vector-like quarks which modifies the $Zb\bar{b}$ couplings in order to provide better agreements with the $A^{0, b}_{\rm FB}$ measurement at LEP~\cite{ALEPH:2005ab}, which favors a positive value for both  $\delta g_{Lb}$ and $\delta g_{Rb}$ (as defined in \autoref{eq:Lzff}).\footnote{See {\it e.g.} Ref~\cite{Gori:2015nqa} for a more updated summary and also future prespectives.  A global fit with the LEP/SLD data shows that the SM predictions of the $Zb\bar{b}$ couplings are just outside the $95\%$ CL region.}  To achieve this, one introduces a vector-like quark doublet, $\Psi_{L,R}$ and a vector-like quark singlet, $\hat{B}_{L,R}$,
\begin{align}
\Psi_{L,R} =&~ \bpm B \\ X \epm   \sim (3,2,-5/6) \,, \nonumber\\
\hat{B}_{L,R}  \sim & ~~  (3,1,-1/3) \,,  \label{eq:BM}
\end{align}
where the three numbers in the bracket denote representations under $SU(3)_c$, $SU(2)_L$, and the $U(1)_Y$ hypercharge, respectively.  Their mass terms and the interactions with SM are given by
\begin{equation}
-\mathcal{L} \supset M_1 \bar{\Psi}_L \Psi_R + M_2 \bar{\hat{B}}_L \hat{B}_R 
+ y_L \bar{Q}_L H \hat{B}_R + y_R \bar{\Psi}_L \tilde{H} b_R +\mbox{h.c.} \,.  \label{eq:Lbm}
\end{equation}
The vev of the Higgs boson generates mixings between the new quarks and the SM ones, which modifies the $Zb\bar{b}$ couplings as 
\begin{equation}
\delta g_{Lb} = \frac{y^2_L v^2}{4M^2_2}  \,, ~~~~~   \delta g_{Rb} =  \frac{y^2_R v^2}{4M^2_1}  \,, \label{eq:y123}
\end{equation}
both are positive as desired.  While \autoref{eq:y123} can be directly derived from the mass mixing, the sum rules in \autoref{eq:sumqh1} provides a transparent connection between the signs of $\delta g_{Lb}$ and $ \delta g_{Rb}$ and the properties of the exotic quarks. 
Taking the 3rd and 4th equations in \autoref{eq:sumqh1}, with an $SU(2)$ rotation one could write
\begin{align}
\frac{4 \, \delta g_{Lb}}{v^2}  = - \frac{2 (c_{Hq} + c'_{Hq})}{\Lambda^2} = \left. \frac{d \tilde{\A}_{t_L \, \phi^-}}{d s} \right|_{s=0}    =&~ \int^\infty_{0}  \frac{ds}{\pi s}  \left( \sigma^{t_L \, \phi^- \to F^{-\frac{1}{3}} } - \sigma^{t_L \, \phi^+ \to F^{\frac{5}{3}} }  \right) + c_\infty \,, \label{eq:sumzbbL}\\%\nonumber\\ 
\frac{4\, \delta g_{Rb}}{v^2} = - \frac{2 c_{Hd}}{\Lambda^2} = \left. \frac{d \tilde{\A}_{b_R \, \phi^-}}{d s} \right|_{s=0}    =&~ \int^\infty_{0}  \frac{ds}{\pi s}  \left( \sigma^{b_R \, \phi^- \to F^{-\frac{4}{3}} } - \sigma^{b_R \, \phi^+ \to F^{\frac{2}{3}} }  \right) + c_\infty \,,   \label{eq:sumzbbR}
\end{align}
where $F$ denotes collectively the BSM fermions with the superscript indicating the electric charge. %
Indeed, $F^{-\frac{1}{3}}$ and $F^{-\frac{4}{3}}$ correspond to $\hat{B}_R$ and $\hat{X}_L$ in \autoref{eq:BM}, while $F^{\frac{5}{3}}$ and $F^{\frac{2}{3}}$ are absent.  The boundary terms $c_\infty$  
are also absent in the BM model.  It is clear from \autoref{eq:sumzbbR} that a charge $-4/3$ quark is required to generate a positive $\delta g_{Rb}$.  A straight forward calculation of the $2\to 1$ cross sections on the right-hand side of \autoref{eq:sumzbbL} and \autoref{eq:sumzbbR} reproduces the results in \autoref{eq:y123}.

\begin{figure}
\centering
\includegraphics[width=0.6\textwidth]{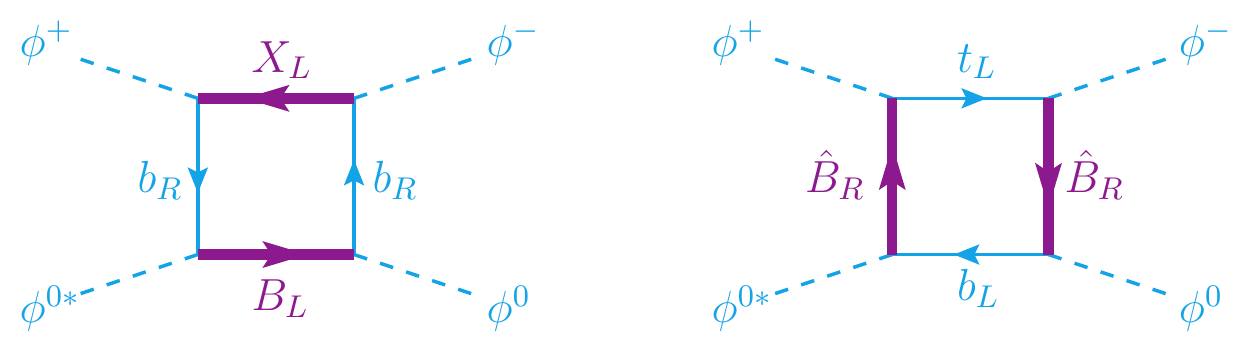}
\caption{The one-loop contributions to $\A^{[2]} (\phi^+ \phi^0 \to  \phi^+ \phi^0)$ from the BM model in \autoref{eq:Lbm} proportional to $y^4_R$ (left) and $y^4_L$ (right).  All external particles are going in.}
\label{fig:bmT}
\end{figure}

A non-zero $T$ parameter is also generated in the BM model.  A direct computation of the fermion loop contributions to the gauge boson propagators gives (assuming $m_b=0$)~\cite{Gori:2015nqa} 
\begin{equation}
T \approx \frac{3}{16\pi^2 \alpha v^2} \left[ \frac{16}{3} \delta g_{Rb}^2 M^2_1 + 4 \delta g_{Lb}^2 M^2_2 - 4\delta g_{Lb} \frac{M^2_2 \, m^2_{\rm top}}{M^2_2-m^2_{\rm top}} \log{\left(\frac{M^2_2}{m^2_{\rm top}}\right)}    \right] \,,  \label{eq:Tsimp}
\end{equation}
in which the first two terms are generated by the fermion loop of the two physical heavy bottom partners while the third term comes from the mixed loop of the mostly $\hat{B}$ partner and top.  Correspondingly, the fermion loops also contribute to $\A^{[2]} (\phi^+ \phi^0 \to  \phi^+ \phi^0)$ in \autoref{eq:a4hO2}.  We note that these contributions must be finite, as otherwise a dimension-6 counter term is needed for the full theory, in contradiction with the full theory being renormalizable.  While a $\phi^+ \phi^0 \phi^- \phi^{0*}$ counter term can be generated at dimension four,  it contributes to $\A^{[0]} (\phi^+ \phi^0 \to  \phi^+ \phi^0)$ rather than $\A^{[2]} (\phi^+ \phi^0 \to  \phi^+ \phi^0)$, and does not have an impact on the sum rule.  In addition, the boundary term $c_\infty$ also vanishes for these contributions as they could only grow as fast as $\log(s)$ for large $s$.  Let us focus on the first two terms in \autoref{eq:Tsimp}, which are proportional to $y^4_R$ and $y^4_L$.  The corresponding loop diagrams of the 4-scalar amplitude are shown in \autoref{fig:bmT}. 
Their contribution can either be computed directly  
or by using the sum rule in \autoref{eq:sumhh2}.  In the latter case, one simply needs to calculate the tree-level $2\to 2$ cross sections of $\phi^+ \phi^{0*} \to \bar{X}_L B_L$ and $\phi^+ \phi^{0*} \to \bar{t_L} b_L$.  We thus obtain
\begin{align}
 \left. \frac{d \tilde{\A}_{\phi^+ \phi^0}}{d s} \right|_{s=0}  =&~   \int^\infty_{0}  \frac{ds}{\pi s} \left( 0 - \sigma^{\phi^+ \phi^{0*} \to \overline{X}_L B_L} - \sigma^{\phi^+ \phi^{0*} \to \overline{t}_L b_L}  \right)  \nonumber\\
=&~ -\frac{y^4_R}{8  \pi^2 \, M^2_1}  -\frac{3\, y^4_L}{32  \pi^2 \, M^2_2} \, .  \label{eq:bmTsum}
\end{align}

\begin{figure}[t]
\centering
\includegraphics[width=0.9\textwidth]{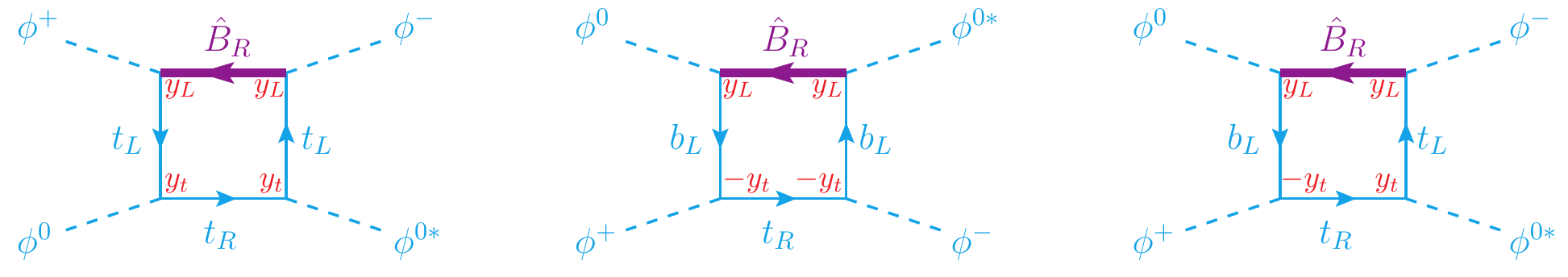}
\caption{The one-loop contributions to $\A^{[2]} (\phi^+ \phi^0 \to  \phi^+ \phi^0)$ from the BM model in \autoref{eq:Lbm} that are proportional to $y^2_t y^2_L$.  The coupling of each vertex (up to some common overall phase) is also labelled.  
All external particles are going in.}
\label{fig:bmT3}
\end{figure}

There are additional 1-loop diagrams contributing to the amplitude $\tilde{\A}_{\phi^+ \phi^0}$ that are proportional to $y^2_t y^2_L$, as shown in \autoref{fig:bmT3}.  
\begin{figure}[t]
\centering
\includegraphics[width=0.4\textwidth]{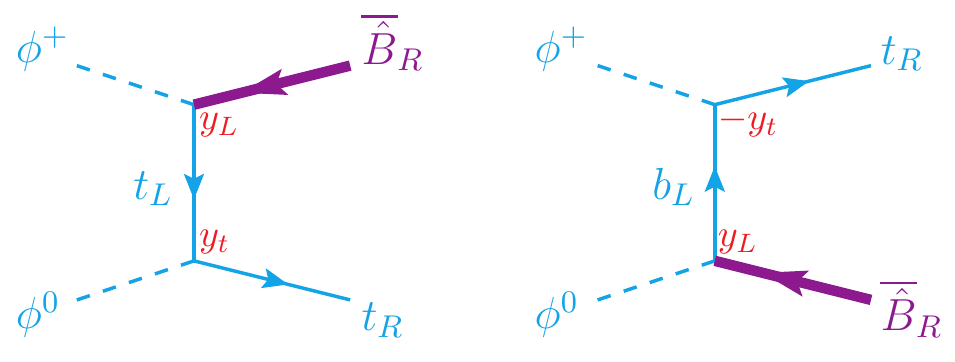} \hspace{2cm}
\includegraphics[width=0.4\textwidth]{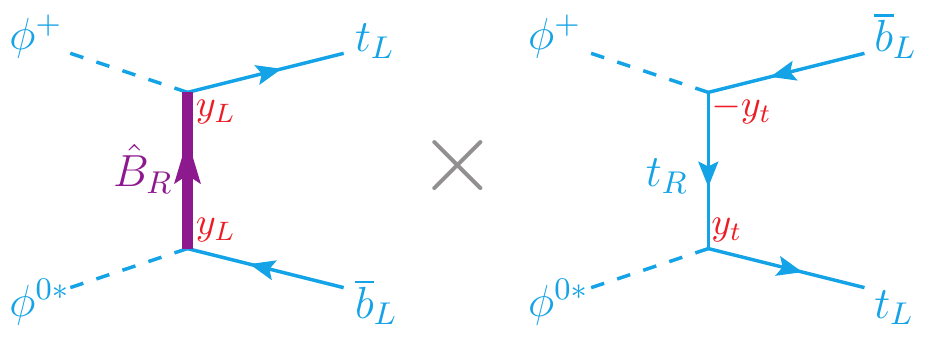}
\caption{ The diagrams for the $2\to2$ cross sections corresponding to the amplitudes in \autoref{fig:bmT3} (via the optical theorem).  The two diagrams on the left contribute to $\sigma(\phi^+ \phi^0 \to \overline{\hat{B}}_R t_R)$.  The two diagrams on the right contribute to $\sigma(\phi^+ \phi^{0*} \to t_L \overline{b}_L)$, but only the interference term is proportional to $y^2_t y^2_L$.}
\label{fig:bmT4}
\end{figure}
The corresponding $2\to 2$ processes are shown in \autoref{fig:bmT4}.  Note in particular that one needs to also include the contribution from the interference term of $\sigma(\phi^+ \phi^{0*} \to t_L b_L)$. 
We also restore a finite $m_t$ while still keeping the scalars massless, which gives
\begin{align}
\left. \frac{d \tilde{\A}_{\phi^+ \phi^0}}{d s} \right|_{s=0}    =&~   \int^\infty_{0}  \frac{ds}{\pi s} \left(  \sigma^{\phi^+ \phi^{0} \to \overline{\hat{B}}_R t_R   } - \sigma^{\phi^+ \phi^{0*} \to \overline{t}_L b_L}  \right)  \nonumber \\
=&~ \frac{ 3 y^2_t y^2_L  }{16\pi^2 M^2_2} \left[ \left( 2\log(\frac{M^2_2}{m^2_t}) -\frac{13}{6}  + ... \right) -   \left( \log(\frac{M^2_2}{m^2_t})  - 2 + ... \right) \right] \nonumber \\
=&~ \frac{3 y^2_t y^2_L }{16\pi^2 \, M^2_2} \left[ \log{(\frac{M^2_2}{m^2_t})} - \frac{1}{6} \right] + \mathcal{O}(\frac{m^2_t}{M^4_2})  \,.  \label{eq:BMmixloop}
\end{align}
Combining \autoref{eq:bmTsum} and the leading log term in \autoref{eq:BMmixloop}, and imposing the relation $\left. \frac{d \tilde{\A}_{\phi^+ \phi^0}}{d s} \right|_{s=0} = - \frac{2 c_T}{\Lambda^2}  = \frac{-2 \alpha T}{v^2}$, we indeed reproduce the result in \autoref{eq:Tsimp}.

From the point of the EFT, we will generate $\mathcal{O}_{Hf}$ operators by integrating out heavy fermions at the tree level at some matching scale close to the heavy fermion masses. On the other hand, the contribution to the low energy $T$ parameter, encapsulated in the SMEFT operator ${\mathcal{O}}_T$, comes from one-loop matching. There is also the contribution from  the operator mixing between  ${\mathcal{O}}_T$ and $\mathcal{O}_{Hf}$ induced by the RGE running from the matching scale to the scale of low energy measurement. 
 To calculate this contribution, we take the RG equation of $c_T$ (for instance, from Ref.~\cite{Jenkins:2013wua}) and keep only the parts proportional to $y^2_t$. %, 
This gives the running of $c_T$ as
\begin{equation}
c_T (\mu) = c_T (\mu_0) - \frac{3y^2_t}{8\pi^2} (-c'_{Hq} + c_{Hu} +c_T) \log (\frac{\mu^2_0}{\mu^2}) \,,  \label{eq:rgbm1}
\end{equation}
where $c_T (\mu_0)$ is the value of $c_T$ evaluated at a reference scale $\mu_0$.  
In the BM model, we have
\begin{equation}
\frac{c'_{Hq}}{\Lambda^2} = -\frac{y^2_L}{4M^2_2} \,, \hspace{1cm}   c_{Hu} = 0\,.
\end{equation}
As $c_T$ itself is generated at one-loop, the $c_T$ coefficient of the log term in \autoref{eq:rgbm1} is formally a two-loop contribution and can be omitted.  We then have
\begin{equation}
\frac{c_T (\mu)}{\Lambda^2} =  \frac{ c_T (\mu_0) }{\Lambda^2} - \frac{3 y^2_t y^2_L}{32 \pi^2 M^2_2} \log (\frac{\mu^2_0}{\mu^2}) \,.
\end{equation}
The running from $\mu_0$ to $\mu$ then generates a contribution to the amplitude
\begin{equation}
\left. \frac{d \tilde{\A}_{\phi^+ \phi^0}}{d s} \right|_{s=0}  = \frac{-2 \left(c_T (\mu) - c_T (\mu_0) \right)}{\Lambda^2} = \frac{3 y^2_t y^2_L}{16 \pi^2 M^2_2} \log (\frac{\mu^2_0}{\mu^2}) \,,
\end{equation}
which, when setting $\mu_0 = M_2$ and $\mu=m_t$, agrees with the log term in \autoref{eq:BMmixloop}.  This is exactly what one would expect, as the RG running of the coupling captures the log enhanced loop contribution to it.  We thus conclude that our direct computation of $c_T$ from the sum rules is consistent with the matching and running procedures of the EFT for the BM model.

%%%%%%%%%%%%%%%%%%%%%%%%%%%%%%%%%%%%%%
\subsection{Models with the $Zb\bar{b}$ custodial symmetry}
\label{sec:zbb}
%%%%%%%%%%%%%%%%%%%%%%

It is also plausible that the discrepancy in the LEP $A^{0, b}_{\rm FB}$ measurement is caused by statistical fluctuations or systematic effects rather than new physics.  In this case, since $t_L$ and $b_L$ are in the same $SU(2)_L$ doublet, the measurement of the $Zb_L\bar{b}_L$ coupling provides very stringent constraints on many new physics models that has extended top sectors.  However, as mentioned in \autoref{sec:sym}, it is possible to impose a symmetry that makes the amplitude in \autoref{eq:sumzbbL} vanish, and protects the $Zb_L\bar{b}_L$ coupling to be SM-like even with the presence of new physics.  
\begin{figure}[t]
\centering
\includegraphics[width=0.55\textwidth]{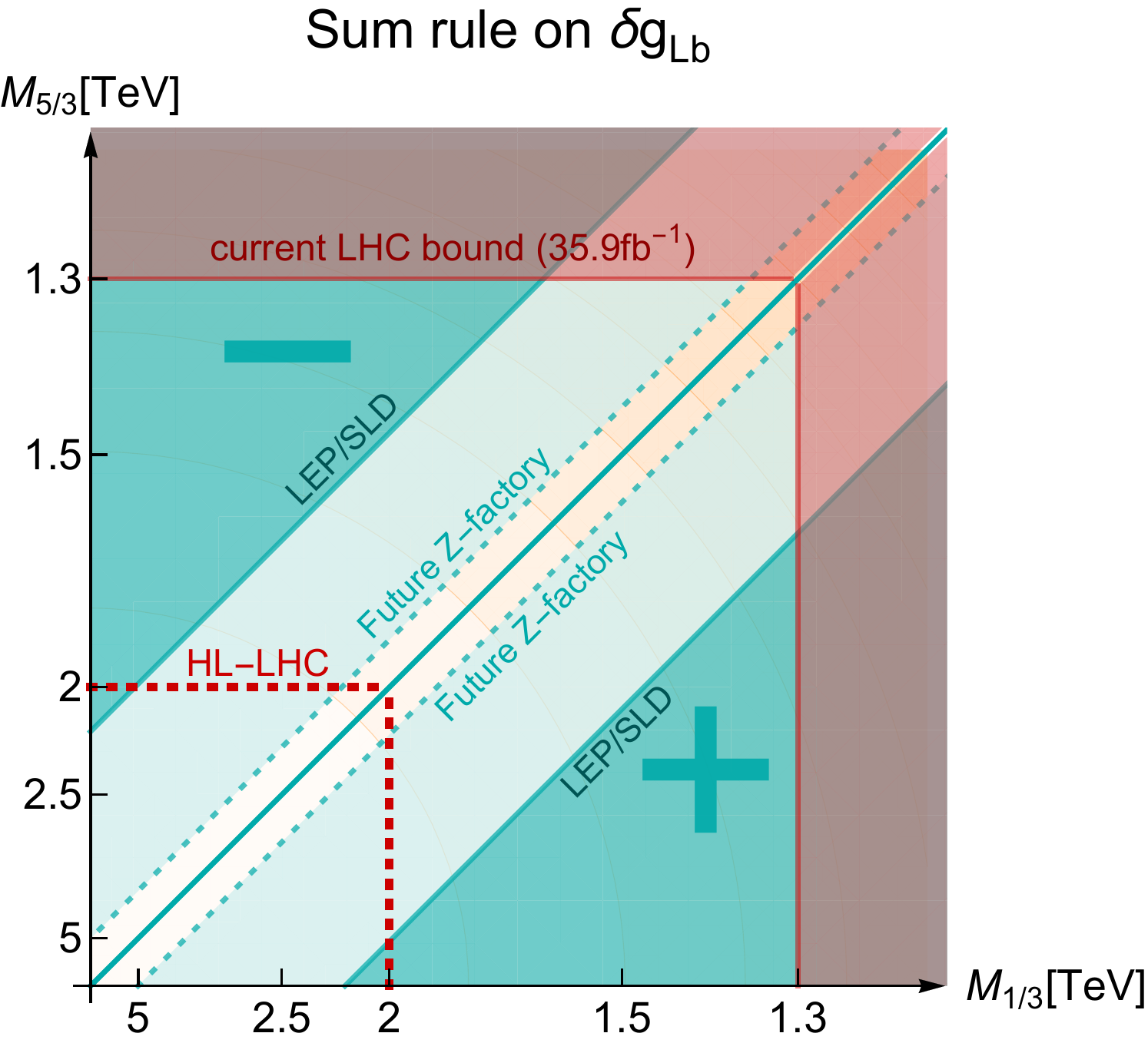}
\caption{
A more specific example of the schematic plot of \autoref{fig:schem} for the $\delta g_{Lb}$ sum rule in \autoref{eq:sumzbbL} with realistic bounds from current and future experiments (all at 95\% CL).  Note that the axes are scaled linearly with $1/M^2$.  The diagonal line corresponds to $\delta g_{Lb}=0$, while the plus (minus) sign denotes the region in which $\delta g_{Lb}$ is positive (negative).  The relevant Yukawa couplings (as in \autoref{eq:Lbm}) are assumed to be one for simplicity.  
}
\label{fig:zbb}
\end{figure}
To illustrate this, we present in \autoref{fig:zbb} the interplay between precision measurements and direct searches for the sum rule in \autoref{eq:sumzbbL}, which is a refined version of  \autoref{fig:schem} with realistic bounds.  For simplicity, we assume the cross section $\sigma^{t_L \, \phi^- \to F^{-\frac{1}{3}} }$ ($\sigma^{t_L \, \phi^+ \to F^{\frac{5}{3}} }$) is generated by a single heavy quark with mass $M_{1/3}$ ($M_{5/3}$), and the relevant Yukawa couplings are set to one.  The constraints are shown in the $(M_{1/3}, M_{5/3})$ plane.   The bounds on $\delta g_{Lb}$ from current and future $Z$-pole measurements are taken from the global fitting results in Ref.~\cite{deBlas:2019wgy}.  The bounds from searches of heavy quarks are taken from Ref.~\cite{Liu:2018hum}.\footnote{We take the bounds from QCD productions which are more robust.  We also assume the bound on the charge $1/3$ quark is similar to the one of the charge $5/3$ quark.}  The bounds from precision measurements are generally more constraining than the ones from direct searches, except for the region near the diagonal line as a result of the sum rule.  This can be realized without tuning model parameters by imposing the symmetry on the amplitude as in \autoref{sec:sym}.  
A common setup in composite Higgs models is to impose a $P_{LR}$ parity in addition to the $SU(2)_L \times SU(2)_R$ symmetries of the Higgs sector, and require that $T^3_L = T^3_R = -1/2$ for $b_L$~\cite{Agashe:2006at}.\footnote{See also Ref.~\cite{Panico:2015jxa} for a recent review on composite Higgs models.}  An important phenomenological consequence of such constructions is the prediction of a heavy exotic quark with electric charge $5/3$, which we have already learned from the sum rule in \autoref{eq:sumzbbL}.

%%%%%%%%%%%%%%%%%%%%%%%%%%%%%%%%%%%%%%%%%%%%%%%%%%%%%%%%%%%%%%%%%%
\section{Conclusion}
\label{sec:con}
%%%%%%%%%%%%%%%%%%%%%%%%%%%%%%%%%%%%%%%%%%%%%%%%%%%%%%%%%%%%%%%%%%

In this paper, we apply the dispersion relations on the forward elastic amplitudes generated by the dimension-6 operators of the Standard Model Effective Field Theory, and derive a set of sum rules on the operator coefficients.  Focusing on the massless limit, we are able to classify and write down the  sum rules using the tool of helicity amplitudes.  
These sum rules offer distinct insights on the connection between the operator coefficients in the EFT and the properties of the full beyond-SM theory.  Their usefulnesses are illustrated in a few benchmark scenarios with scalar and fermionic extensions of the SM.  As an application, the sum rules also help us identify the possible symmetries that suppress the contributions of dimension-6 operators in certain amplitudes, which can be connected to the custodial symmetries that protects the $T$ parameter or the fermion gauge couplings.

It is somewhat unsatisfying that our sum rules only cover a subset of the dimension-6 operator coefficients.  While the forward elastic amplitudes give the most straight forward sum rules, recent studies have also found interesting implications for the SMEFT dimension-6 operators from non-forward amplitudes~\cite{Remmen:2020uze}.   It is desirable, if possible, to obtain meaningful sum rules also for inelastic amplitudes.  A novel approach, based on convex geometries, is studied in Ref.~\cite{Bellazzini:2014waa} and recently revisited in Ref.~\cite{Zhang:2020jyn} for systematically obtaining the positivity bounds on dimension-8 operator coefficients.  
This approach however may not be directly applicable to the sum rules of dimension-6 operators.  A more general question, sometimes named as the {\it Inverse Problem}~\cite{ArkaniHamed:2005px, Dawson:2020oco}, can be phrased as follows: Given the measured values of the operator coefficients around the electroweak scale, to what extent can we possibly determine the nature of the new physics beyond the SM?  New developments on the amplitude tools might be able to help us further tackle this problem in the future.

%%%%%%%%%%%%%%%%%%%%%%%%%%%%%%%%%%%%%%%%%%%%%%%%%%%%%%%%%%%%%%%%%%
\subsection*{Acknowledgments}
We thank Cen Zhang for many useful discussions.  
The work of JG has been supported by the Cluster of Excellence ``Precision Physics, Fundamental Interactions, and Structure of Matter'' (PRISMA+ EXC 2118/1) funded by the German Research Foundation (DFG) within the German Excellence Strategy (Project ID 39083149). LTW is supported by the DOE grant de-sc0013642. 

%%%%%%%%%%%%%%%%%%%%%%%%%%%%%%%%%%%%%%%%%%%%%%%%%%%%%%%%%%%%%%%%%%

\appendix

%%%%%%%%%%%%%%%%%%%%%%%%%%%%%%%
\section{Essential results of the spinor helicity formalism}
\label{app:heli}
%%%%%%%%%%%%%%%%%%%%%%%%%%%%%%%

Here we try to provide a minimal set of results of the spinor helicity formalism that are needed for our analysis in \autoref{sec:osamp}, with many details omitted.  We refer the readers to some of the recent reviews ({\it e.g.} Refs.~\cite{Elvang:2013cua, Dixon:2013uaa, Cheung:2017pzi}) for a more complete introduction of the subject.  We work with the mostly negative-metric convention, $\eta_{\mu\nu} = {\rm diag}(+1,-1,-1,-1)$, and assume all particles are massless. 

We start by writing a 4-momentum in the bi-spinor forms
\begin{equation}
p_{\alpha\dot{\alpha}} \equiv p_\mu (\sigma^\mu)_{\alpha\dot{\alpha}}   \,,  \hspace{1cm}
p^{\dot{\alpha}\alpha} \equiv p_\mu (\bar{\sigma}^\mu)^{\dot{\alpha}\alpha}  \,,
\end{equation}
where $\sigma^\mu = (1, \sigma^i)$ and $\bar{\sigma}^\mu = (1, -\sigma^i)$, and $\sigma^{1,2,3}$ are the Pauli matrices.  $p_{\alpha\dot{\alpha}}$ and $p^{\dot{\alpha}\alpha}$ are $2\times2$ matrices, and their determinants are given by
\begin{equation}
{\rm det}(p) = p^\mu p_\mu = m^2 = 0 \,.
\end{equation}
Thus, in the massless limit, $p_{\alpha\dot{\alpha}}$ and $p^{\alpha\dot{\alpha}}$ have rank 1, and can be written as products of a pair of 2-component spinors,
\begin{equation}
p_{\alpha\dot{\alpha}} = \lambda_\alpha \tilde{\lambda}_{\dot{\alpha}} \equiv | p \ra [p |  \,,  \hspace{1cm}
p^{\dot{\alpha}\alpha}  =  \tilde{\lambda}^{\dot{\alpha}} \lambda^\alpha \equiv |p] \la p | \,, 
\end{equation}
where we have introduced the half-brackets as shorthands for the spinors ($\lambda_\alpha \to | p \ra$, $ \tilde{\lambda}^{\dot{\alpha}} \to |p]$).    
Note that the choices of spinors are not unique -- a simultaneous scaling of the form (denoted as the little group scaling)
\begin{equation}
|p\ra \to t |p\ra  \,,  \hspace{1cm}
|p]  \to t^{-1} |p]   \,,  \label{eq:lgs}
\end{equation}
leaves $p_{\alpha\dot{\alpha}}$ invariant.  For real momenta, $p_{\alpha\dot{\alpha}}$ is Hermitian, implying that $[p| = (|p\ra)^*$, and the $t$ in \autoref{eq:lgs} can only be a phase.  For complex momenta, $|p]$ and $|p\ra$ can be treated as independent quantities.

$|p]$ and $|p\ra$ transform differently under the Lorentz group. In fact, they can be associated with the helicity of the particle, with $|p]$ ($|p\ra$) corresponding to helicity $+1/2$ ($-1/2$).  This imposes strong constraints on the form of amplitudes that can be written down.  In particular, a $n$-point amplitude of particles with helicities $h_{1, \,...\,,n}$ is little group covariant with weight
\begin{equation}
\A(1^{h_1} ,\,...\,, n^{h_n}) \to \underset{i}{\prod} \, t^{-2h_i}_i \, \A(1^{h_1} ,\,...\,, n^{h_n}) \,,  \label{eq:Algs}
\end{equation}
which is essential in fixing the forms of the amplitudes in \autoref{tab:a4elastic}.

Lorentz invariant quantities can be constructed by contracting the indices of two $\lambda$s (or two $\tilde{\lambda}$s) using the antisymmetric Levi-Civitas symbol.   This can be conveniently written in terms of angle or square brackets as
\begin{equation}
 \la i \, j \ra \equiv \epsilon^{\alpha\beta} \lambda_{i\alpha}   \lambda_{j\beta} =   \lambda_i^{~\alpha}   \lambda_{j\alpha}  \,,  \hspace{1cm}
 [i \, j]   \equiv  \epsilon^{\dot{\alpha}\dot{\beta}} \tilde{\lambda}_{i\dot{\alpha}}   \tilde{\lambda}_{j\dot{\beta}} =  \tilde{\lambda}_{i\dot{\alpha}}   \tilde{\lambda}_j^{~\dot{\alpha}} \,,
\end{equation}
where we have further introduced the shorthand $|p_i\ra \to | i \ra$ and so on.  Note also that $\la i \, i \ra =  [i \, i]=0$ due to their antisymmetric nature.  For 4-point amplitudes, one could explicitly work out the Mandelstam variables, which turn out to be (in the massless limit)
\begin{align}
s\equiv s_{12} =&~ (p_1+p_2)^2 = 2p_1p_2 =  \la 1 2 \ra [1 2] =  \la 3 4 \ra [3 4]  \,, \nonumber \\
t\equiv s_{13} =&~ (p_1+p_3)^2 = 2p_1p_3 = \la 1 3 \ra [1 3] = \la 2 4 \ra [2 4]  \,,  \nonumber \\  
u\equiv s_{14} =&~ (p_1+p_4)^2 = 2p_1p_4 = \la 1 4 \ra [1 4] = \la 2 3 \ra [2 3] \,.  
\end{align}
It is also useful to relate different spinor products using momentum conservation. In particular, we could write
\begin{equation}
\la 1 2\ra[23] = - \la 1 1\ra[13] - \la 1 3\ra[33] - \la 1 4\ra[43] = - \la 1 4\ra[43] \,,  \label{eq:a1223}
\end{equation}
where we have used $p_2 = -p_1 -p_3-p_4$ and $\la 1 1\ra = [33] =0$.  Thus, we see in \autoref{tab:a4elastic} that the spinor form of $\A^{[2]}_4$ for $\A(\psi^- \phi \psi^+ \phi^*)$ only has one independent term, which we choose to be $\la 1 2\ra[23]$.

One important feature of an on-shell 3-point amplitude is that it is composed of either only angle brackets or only square brackets.\footnote{We shall omit the derivation here, which can be found in Refs.~\cite{Elvang:2013cua, Dixon:2013uaa, Cheung:2017pzi}.  Also note that for massless particles, the 3-point amplitudes vanish for real momenta, but they can be written down for complex momenta.} Its form is thus fixed by the little group scaling, given by
\begin{equation}
\A(1^{h_1} 2^{h_2} 3^{h_3}) = \left\{ \begin{matrix} ~g \, \la 12\ra^{h_3 -h_1 -h_2}  \la 23\ra^{h_1 -h_2 -h_3}  \la 31\ra^{h_2 -h_3 -h_1}  \,, ~~~ h_1+h_2+h_3 \leq 0 \\ g \, [12]^{h_1 +h_2 -h_3}  [ 23 ]^{h_2 +h_3 -h_1}    [ 31 ]^{h_3 + h_1 -h_2}  \,, ~~~~ h_1+h_2+h_3 \geq 0  \end{matrix}  \right. \,,
\end{equation}
where $g$ is the coupling associated with the 3-point vertex.  Assuming $h_1+h_2+h_3>0$, the total dimension of the amplitude is given by
\begin{equation}
[\A] = [g]+ h_1+h_2+h_3 = 1 \,, \label{eq:DA3}
\end{equation}
where in the last step we used the fact that the dimension of a $n$-point amplitude is $4-n$.  Considering the case that $[g]=-2$, {\it i.e.} the 3-point amplitude is generated by a dimension-6 operator, and assuming $|h|\leq 1$ (considering particles with spin less or equal to one), we see that the only solution to \autoref{eq:DA3} is
\begin{equation}
h_1 = h_2 = h_3 =1 \,,
\end{equation}
which corresponds to 3 vectors with the same helicity, $\A(V^+V^+V^+)$.  Similarly, for $h_1+h_2+h_3<0$ we find that the only possible amplitude with $[g]=-2$ is  $\A(V^-V^-V^-)$.  We also see that the dimension of $g$ could not be smaller than $-2$, suggesting that the 3-point massless on-shell amplitude could not be generated by operators of dimension-8 or higher (assuming $|h|\leq 1$).

%%%%%%%%%%%%%%%%%%%%%%%%%%%%%%%%%%%%%%%%%%%%%%%%%%%%%%%%%%%%%%%%%%%
\section{The forward limit}
\label{sec:afor}
%%%%%%%%%%%%%%%%%%%%%%%%%%%%%%%%%%%%%%%%%%%%%%%%%%%%%

Here we take the amplitudes in \autoref{tab:a4elastic} and derive their forms in the forward limit.  This is straightforward for the all-scalar amplitudes --- they can be written in terms of the Mandelstam variables, and we simply set $t =0$ and use the massless relation $s+t+u =0$ to write $u = -s$.  
For amplitudes involving spins, this is slightly more complicated as they are covariant under the little groups transformations of the external particles.  Remarkably, for massless particles with any spins, it is shown (in the helicity basis) that the elastic amplitudes in the forward limit are always invariant under the little group scaling, and can be treated as if they are scalar amplitudes~\cite{Bellazzini:2016xrt}.  Here we try to provide a somewhat simpler derivation within the framework of the helicity amplitudes.  
The key observation is that in a forward elastic scattering, by definition, the incoming particle 1 and the outgoing particle 3 (or 2 and 4) are the same particle with the same momentum and quantum numbers.  In general, each particle has a different scaling and the amplitude has to transform according to the helicities of the external particles, as shown in \autoref{eq:Algs}.  However, in the forward elastic limit, one could impose without the loss of generality that (note the $t$ here is not the Mandelstam variable $t$)
\begin{equation}
 t_1 = t_3  \,, \hspace{1cm}   t_2 = t_4 \,. \label{eq:lgf}
\end{equation}
Physically, the little group scaling of massless particles corresponds to the rotation around the axis of the momentum and the translation along it.  \autoref{eq:lgf} is indeed only possible in the forward limit, where particle 1 and 3 are along the same direction.

We recall that in the all-in/all-out convention, particle 1 and 3 (2 and 4) have opposite helicities, $h_1 = -h_3$,  $h_2 = -h_4$, as suggested in \autoref{tab:a4elastic}.  Under \autoref{eq:lgf}, the amplitude in the forward limit (denoted as $\tilde{\A}$) thus scales as
\begin{equation}
\tilde{\A} \sim \underset{i}{\Pi}\, t_i^{-2h_i} = t_1^{-2(h_1+h_3)} t_2^{-2(h_2+h_4)} = 1 \,, \label{eq:ainv}
\end{equation}
which indeed shows that forward elastic amplitudes are invariant under little group scaling in the massless case, regardless of the particle spins.  As such, they can be written in terms of the Mandelstam variables, and more specifically, in terms of $s$ alone by setting $t=0$ and $u=-s$.    
For the terms in \autoref{tab:a4elastic}, we then have
\begin{equation}
\tilde{\A}^{[2]}_4 \equiv \A^{[2]}_4 |_{t\to0} \propto s  \,, \hspace{1cm}  \tilde{\A}^{[4]}_4 \equiv  \A^{[4]}_4 |_{t\to0} \propto s^2 \,.  \label{eq:foamps2}
\end{equation}
To verify  \autoref{eq:foamps2}, 
we note that the Mandelstam variables are invariant under the simultaneous exchanges $1 \leftrightarrow 3$ and $2 \leftrightarrow 4$, and so should the forward amplitudes, 
\begin{align}
x \equiv&~  \la 1 2\ra[23] \left|_{t\to 0} \right. =    \la 34 \ra [ 4 1] \left|_{t\to 0} \right. =    -  \la 32 \ra [  21 ]  \left|_{t\to 0} \right.  \,,  \nonumber\\ 
  y \equiv&~  \la 1 2\ra[34]\left|_{t\to 0} \right. =   \la 34\ra [12] \left|_{t\to 0} \right. \,, \end{align}
where the shorthands $x$ and $y$ are defined purely for convenience.  We then have
\begin{align}
x^2 =&~ - \la 1 2\ra [12]   \la 2 3\ra [23]  \left|_{t\to 0} \right. = - s u \left|_{t\to 0} \right. =   s^2 \,,  \nonumber\\
 y^2 =&~ \la 1 2\ra [12]   \la 34 \ra [34]\left|_{t\to 0} \right. = s^2  \,.
\end{align}
Thus, without loss of generality one could pick up the solution with a positive sign and write 
\begin{align}
\la 1 2\ra[23]  \underset{t=0}{\rightarrow}&~   s \,,  \nonumber  \\
\la 1 2\ra[34]  \underset{t=0}{\rightarrow}&~   s  \,.   \label{eq:foamp}
\end{align}
We have also shown in \autoref{sec:eaeft} that the only pole generated by higher dimensional operators in the 4-point elastic amplitudes is the $s$-channel pole in the $\A^{[4]}_4$ of the $V^-_1 V^-_2 V^+_1 V^+_2$ amplitude in \autoref{tab:a4elastic}.  Without $t$-channel poles, all amplitudes are finite in the forward limit.  Therefore, the amplitude expansion in \autoref{eq:ampexp} can be written in the forward limit as
\begin{equation}
\tilde{\A}_4 = \sum_n \tilde{g}_{[-2n]} s^n \,,   \label{eq:ampexp2}
\end{equation}
where $\tilde{g}$ are the coefficients of each term with mass dimension $-2n$ as labelled in the subscript.  Certain SM contribution, such as a $t$-channel photon exchange, could invalidate \autoref{eq:ampexp2}, but is known and can be subtracted, as mentioned in \autoref{sec:valid}.  \autoref{eq:ampexp2} now matches exactly with the expansion in \autoref{eq:A02} in the limit $\mu^2\to 0$, with $ \tilde{g}_{[-2n]} = c_n$.  Each term in the amplitude expansion ($\A^{[2]}_4$, $\A^{[4]}_4$, ...) thus provides one sum rule for operator coefficients of the corresponding dimension (6, 8, ...).

%%%%%%%%%%%%%%%%%%%%%%%%%%%%%%%%%%%%%%%%%%%%%%%%%%%%%%%%%%%%%%%%%%

\bibliographystyle{JHEP}
\bibliography{pos}

\end{document}